\documentclass[11pt,a4paper]{article}
\usepackage[utf8]{inputenc}
\usepackage{fullpage}
\usepackage{setspace} 
\usepackage{indentfirst}
\usepackage[usenames,dvipsnames]{color}
\usepackage{import}
\usepackage{amsmath}
\usepackage{amssymb}
\usepackage{latexsym}

\usepackage{listings}
\usepackage{bm,bbm}
\usepackage{url}
\usepackage{float} 

\usepackage{ifpdf}
\newif\ifpdf
\ifx\pdfoutput\undefined
\pdffalse % we are not running PDFLaTeX
\else
\pdfoutput=1 % we are running PDFLaTeX
\pdftrue
\fi
\ifpdf
\usepackage[pdftex]{graphicx}
\graphicspath{{}} \DeclareGraphicsExtensions{.pdf}
\else
\usepackage{graphicx}
\graphicspath{{}} \DeclareGraphicsExtensions{.eps}
\fi
\usepackage{color}
\usepackage{boxedminipage} % Surround parts of graphics with box
\usepackage{tikz}
\usetikzlibrary{shapes,snakes,arrows,patterns}
\usepackage{subfigure}

\newcommand{\ie}{\emph{i.e.~}}

\newcommand\B{\mathcal B}

\newcommand\Surf{\mathcal S}
\newcommand\TensorTwoD{\underline{\underline{\sigma}}}

\newcommand\TensorThreeD{\vec{\vec{\sigma}}}

\newcommand\V{\mathcal{V}}

\def\div{\mathop{\rm div}\nolimits} 
\newcommand{\dsp}{\displaystyle}

\newcommand{\vm}{V_m}
\newcommand{\ue}{u_e}
\newcommand{\ui}{u_i}

\definecolor{myblue}{rgb}{0.0,0.57,0.81}
\definecolor{myblueheart}{rgb}{0.082,0.48,0.764}
\definecolor{myred}{rgb}{0.86,0,0.17}
\definecolor{myredheart}{rgb}{0.99,0.12,0.196}
\definecolor{mygreen}{rgb}{0,0.58,0}
\definecolor{mygray}{rgb}{0.4,0.4,0.4}
\definecolor{myyellow}{rgb}{1,0.84,0.024}
\definecolor{mycyan}{rgb}{0.19,0.835,0.87}
\definecolor{mypurple}{rgb}{0.635,0.055,0.67}
\definecolor{myorange}{rgb}{0.86,0.44,0.145}

\usepackage[normalem]{ulem}
\usepackage{marvosym}

\newcommand{\modify}[2]{{\color{red}\sout{#1}$\mapsto$}{\color{blue}#2}}

\begin{document}

\title{Numerical simulation of electrocardiograms for full cardiac cycles in healthy and pathological conditions}

%\date{June 2015}

\author{Elisa Schenone\footnote{Sorbonne Universit\'es UPMC \& Inria Paris-Rocquencourt, France}, 
Annabelle Collin
\footnote{Inria Saclay Ile-de-France, France},
Jean-Fr\'ed\'eric Gerbeau
\footnote{Inria Paris-Rocquencourt \& Sorbonne Universit\'es UPMC, France}
}

\maketitle

\begin{abstract}
This work is dedicated to the simulation of full cycles of the electrical activity of the heart and the corresponding body surface potential. The model is based on a realistic torso and heart anatomy, including ventricles and atria. One of the specificities of our approach is to model the atria as a surface, which is the kind of data typically provided by medical imaging for thin volumes. The bidomain equations are considered in their usual formulation in the ventricles, and in a surface formulation on the atria. Two ionic models are used: the Courtemanche-Ramirez-Nattel model on the atria, and the ``Minimal model for human Ventricular action potentials'' (MV) by Bueno-Orovio,  Cherry and Fenton in the ventricles. The heart is weakly coupled to the torso by a Robin boundary condition based on a resistor-capacitor transmission condition. Various ECGs are simulated in healthy and pathological conditions (left and right bundle branch blocks, Bachmann's bundle block,  Wolff-Parkinson-White syndrome). To assess the numerical ECGs, we use several qualitative and quantitative criteria found in the medical literature. Our simulator can also be used to generate the signals measured by a vest of electrodes. This capability is illustrated at the end of the article. 
\end{abstract}

\noindent{\bf Keywords:} Electrocardiograms; bidomain model; atria; ventricles; electrode vest;

% \noindent{\bf Mathematics Subject Classification (2010):} 

%=========================================================================
\section*{Introduction}

An electrocardiogram (ECG) is a recording of the electrical activity of the heart~\cite{malmivuo-plonsey-95,wartak-75}. The standard 12-lead ECG is obtained from 9 electrodes located on the body surface. This non-invasive and inexpensive procedure is probably the most used clinical tool for the detection of cardiac pathologies. The motivation for the modeling and the simulation of ECGs is twofold. First, the ECG is a simple output of the complex simulation of the cardiac electrical activity: while the latter is difficult to validate without invasive measurements, the former is easy to assess by a medical expert. Second, the simulation of ECGs, also known as the forward problem of electrocardiography~\cite{lines-buist-grottum-03}, can be viewed as a step toward the inverse problem of electrocardiography~\cite{jiang-ling-crozier-08,liu-liu-he-06,nielsen-cai-lysaker-07,wang-macleod-johnson-13}. Indeed, the inverse problem can be reformulated as a problem of identifying the parameters of the model used to simulate ECGs. From this perspective, the design of a model based on biophysical principles and able to produce ECGs in healthy and pathological condition is an important endeavor. This is the main purpose of this study. 

Many attempts at simulating ECGs can be found in the literature \cite{bishop-plank-11,huiskamp-98,lines-buist-grottum-03,keller-davis-seemann-07}. The simulation of 12-lead ECG based on partial differential equations (PDE) -- as opposed to cellular automata~\cite{wei1-okazaki-hosaka-95} -- appeared during the last decade~\cite{boulakia-cazeau-fernandez-10,martin-drochon-gerbeau-12,potse-dube-gulrajani-03,potse-dube-vinet-09,trudel-dube-potse-04}. More recently, a focus on the T-wave was proposed~\cite{hurtado-kuhl-14,keller-weiss-seemann-12}. Simulations of ECGs of patients with Left Bundle Branch Blocks were obtained~\cite{potse-krause-auricchio-14}.
 The precordials leads of the ECG of a rabbit was approximated with a variant of the concept of cardiac vector~\cite{krishnamoorthi-klug-2014}. Because of the difficulty in imaging and modeling the atria,  all these studies only consider the ventricles. As a consequence, they cannot produce the P-wave of the ECG.  On the contrary, papers considering atria usually do not include ventricles \cite{krueger-seemann-dossel-13} or do not provide ECGs~\cite{harrild-graig-00,krueger-schmidt-tobon-11}.

Electrically, the atria and the ventricles are coupled only in a very small area, the atrioventricular node. This explains why these studies can concentrate on the upper and lower chambers of the heart only. Nevertheless, to produce full ECGs, or to simulate pathologies like Wolff-Parkinson-White syndrome, it is necessary to address both atria and ventricles. Recently, a full cycle electrocardiogram was proposed for an idealized geometry~\cite{solvilj-magjarevic-lovell-13}. To our knowledge, the present paper is the first one to propose a full cycle ECG -- including the P, QRS and T waves -- based on a real anatomy of the heart and on the simulation of a PDE-based biophysical model. 

Another limitation of the existing works on ECG simulation is their lack of precise evaluation of the results. This question is indeed delicate since a ``healthy ECG'' is not a unique object, and there is no obvious metric to measure the discrepancy between this somehow fuzzy notion and the result of a simulation. In this paper, we gather many criteria, found in the medical literature, that can be used to assess both qualitatively and quantitatively the numerical ECGs. We show that our healthy ECG fulfills almost all the qualitative and quantitative properties of real ECGs, which is a significant progress with respect to the state of art. Pathological cases are also investigated in order to show the capability of our model to predict the features used by medical doctors to detect a disease. 

Here is a brief description of our approach. A standard $3$D bidomain model is used for the ventricles \cite{tung-78,sachse-04} and a recent asymptotic surface-based bidomain model is used for the atria~\cite{chapelle-collin-gerbeau-12,collin-gerbeau-hocini-13}. Two different ionic models are considered: the Courtemanche-Ramirez-Nattel model \cite{courtemanche-ramirez-nattel-98} on the atria, and the ``Minimal model for human Ventricular action potentials''~(MV) in the ventricles~\cite{buenoOrovio-cherry-fenton-08}. The coupling between the heart and the body is based on a resistor-capacitor coupling condition~\cite{boulakia-cazeau-fernandez-10}. 
 
The outline of this article is as follows. In Section~1, the geometry of the heart used for the simulation is described. It is based on a surface region for the atria and a volume region for the ventricles. Its main characteristics are compared with those of a normal human heart. Section~2 deals with the biophysical modeling of the atria and the ventricles and the coupling condition with the rest of the body. Section~3 concerns the simulations of the standard 12-lead ECG. A healthy case is given and validated against numerous criteria used to assess real electrocardiograms. Some pathological cases are also studied: left and right bundle branch blocks, Bachmann's bundle block, and the Wolff-Parkinson-White syndrome which is a pathology caused by the presence of an abnormal accessory electrical conduction pathway between the atria and the ventricles. In the last part of this section, we investigate the impact of the ionic models on the ECGs by using in the ventricle and in the atria the phenomenological Mitchell-Schaeffer model~\cite{mitchell-schaeffer-03}. In Section 4, we show that our simulator can also produce signals that are richer than the standard ECG. As an illustration, we analyse the potential measured by a ``virtual electrode vest'' made of 1216 electrodes. In particular, the correlation between the signals of different electrodes is studied. We also give an analysis of the dependence of the electrode measures with respect to their positions on the body.

\section{Whole heart mesh}  \label{sec:mesh}
To obtain full cycle ECGs, the first step is to build a whole heart realistic mesh. The ventricles can be easily obtained from medical imaging and meshed in 3D. On the contrary, the atria have a very thin wall which makes them difficult to image in 3D. In addition, generating a 3D mesh on these very thin volumes would dramatically, and uselessly, increase the computational cost. For these reasons, we choose to model the geometry of the atria as a surface. We therefore obtain an hybrid mesh, made of tetrahedra in the ventricles and of triangles in the atria. 

\begin{figure}
	\centering
	\includegraphics[width=\textwidth]{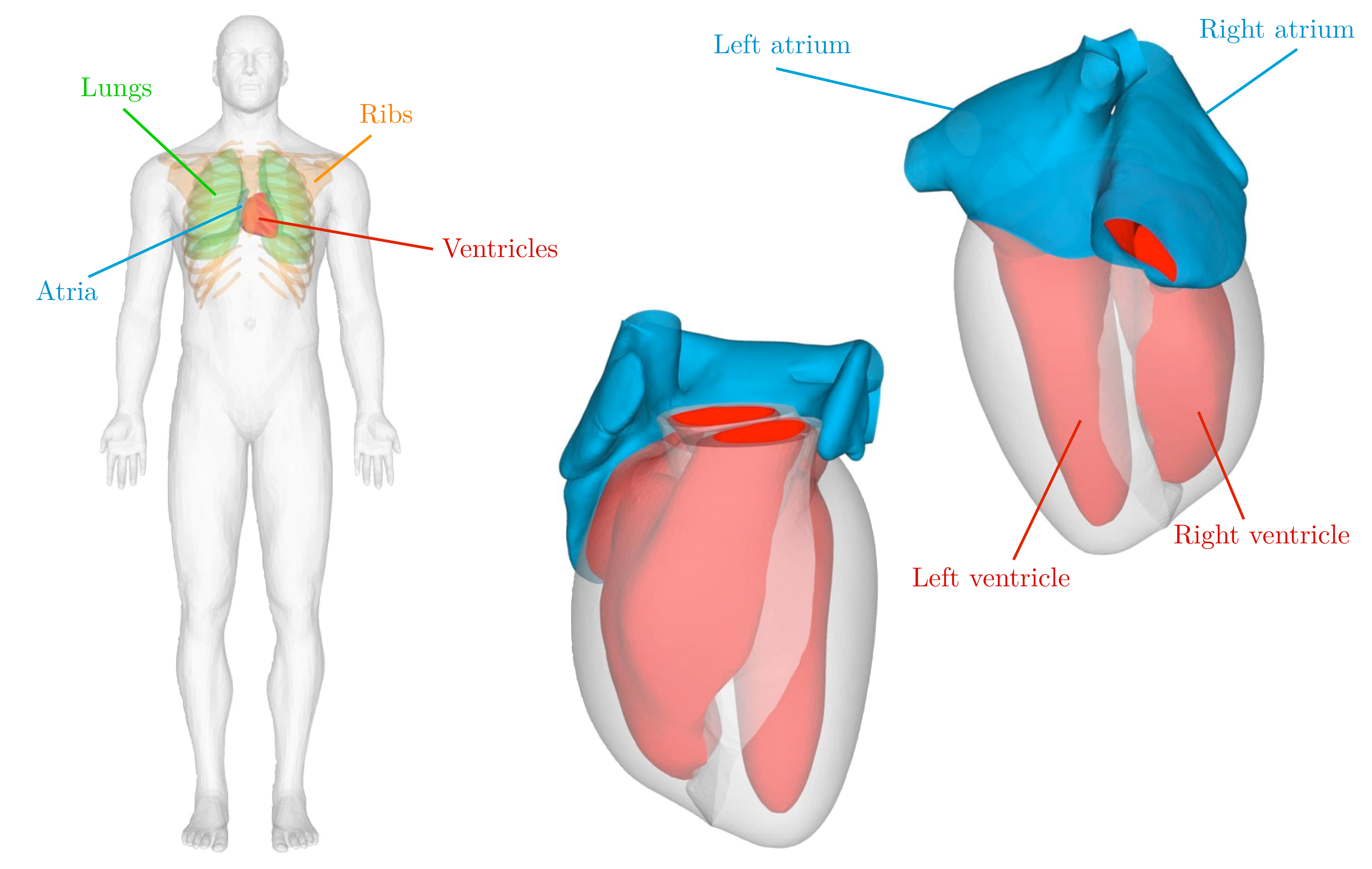}
	\caption{Whole heart mesh (right) and body mesh (left).}
	\label{fig:mesh}
\end{figure}

The heart model was obtained from an anatomical data set called \emph{Zygote}\footnote{\url{www.3dscience.com}}. The \texttt{3-matic} software was used to obtain a surface mesh satisfying the standard quality criteria of a finite element mesh, and \texttt{Yams}~\cite{yamsRT0252} to refine the surface mesh. Then, the volume of the two ventricles was meshed using \texttt{Gmsh}~\cite{gmsh-09}. We can see in Figure~\ref{fig:mesh} different views of the whole mesh, which contains about 230,600 tetrahedra, 73,500 triangles and 67,300 vertices (note that we checked that the ECG simulation was not affected by refining the mesh in the ventricles, increasing the number of tetrahedra up to 2,511,400).
A simplified mesh of the body (Fig.~\ref{fig:mesh}), including the lungs and the ribs, was also built from the \emph{Zygote} data set and the aforementioned software. The body mesh contains 408,171 tetrahedra, 89,222 triangles and 85,196 vertices.

The mechanical deformation of the heart is not taken into account in this work. The dimensions of the fixed domain correspond to the end of the systole (small ventricles, large atria). Table~\ref{tab:HeartMeas} shows a comparison of a few dimensions of the geometrical model with standard end-systolic values. The following quantities are compared: left ventricle volume and mass, mitral and aortic valves diameters, left atrium major axis, area, volume and four pulmonary veins diameters. We observe a good agreement with the values found in the literature  \cite{chatterjee-massie-07,hudsmith-petersen-francis-05,ormiston-shah-tei-82}.  We also have a good agreement for the diameters of mitral \cite{ennis-ruddbarnard-li-10} and aortic \cite{zwink-burwash-miyake-94} valves, the surface of the left atrium~\cite{cohen-white-sochowski-95,jiamsripong-honda-reuss-08}.

\begin{table}
\centering
\begin{tabular}{ccccc}
\hline\hline
Left Ventricle           & Volume (ml) & Mass (g) & Mitral (cm) & Aortic (cm) \\ 
\hline
\hline
\textit{Measures}		 & $53.7$ & $111.3$  & $2.9$ & $2.3$ \\

\textit{Reference} & $46\pm11$ & $112\pm27$ & $2.5\pm0.4$ & $2.3\pm0.2$ \\
\hline\hline
Left Atrium				& Major axis (cm) & Area (cm$\mbox{}^2$) & Volume (ml) & Pulmonary Veins (cm) \\
\hline\hline
\textit{Measures}		& $4.65$ & $15.9$ & $47$ & $1.14-1.45$ \\
\textit{Reference}		& $3.4\pm0.6$ & $17.5\pm2.5$ & $58\pm34$ & $1.3\pm0.2$ \\
\hline
\end{tabular}
\caption{Comparisons of the model dimensions with typical end-systolic values found in the literature \cite{chatterjee-massie-07,hudsmith-petersen-francis-05,ormiston-shah-tei-82,ennis-ruddbarnard-li-10,zwink-burwash-miyake-94,cohen-white-sochowski-95,jiamsripong-honda-reuss-08}}
\label{tab:HeartMeas}
\end{table}

\begin{figure}
	\centering
	\includegraphics[width=\textwidth]{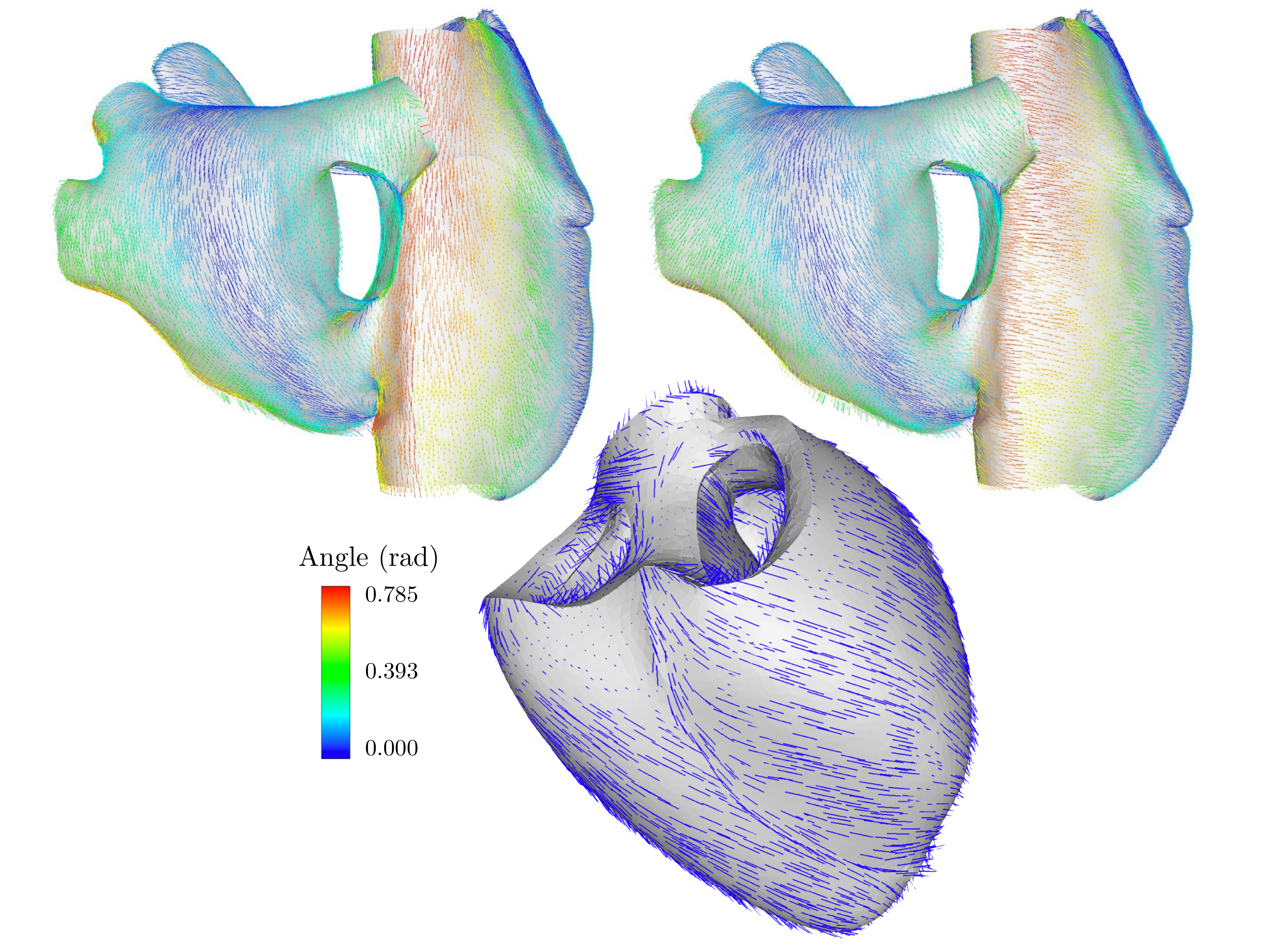}
	\caption{Fibers directions at the atrial endocardium (top-left) and atrial epicardium (top-right), and in ventricles (bottom).}
	\label{fig:fibers}
\end{figure}

Cardiac tissue has a fiber architecture. The electrical conductivity is higher along the fibers than in the transverse direction. This implies that the fiber orientation is very important in the study of the electrical activity of the heart. To identify and to prescribe the fibers at the endocardium and at the epicardium of the atria, we used~\cite{ho-anderson-sanchezQuintana-02,ho-sanchezQuintana-09,krueger-schmidt-tobon-11}. As we can see in Figure~\ref{fig:fibers}~(top), the fibers orientation may vary extremely quickly across the thickness. The colors represent the angle $\theta$ defined as half of the angular difference between the endocardium and the epicardium. We used \cite{nash-hunter-00,streeter-79} to prescribe the fibers in the ventricles, see Figure~\ref{fig:fibers}~(bottom).

\begin{figure}
	\centering
	\includegraphics[width=1.0\textwidth]{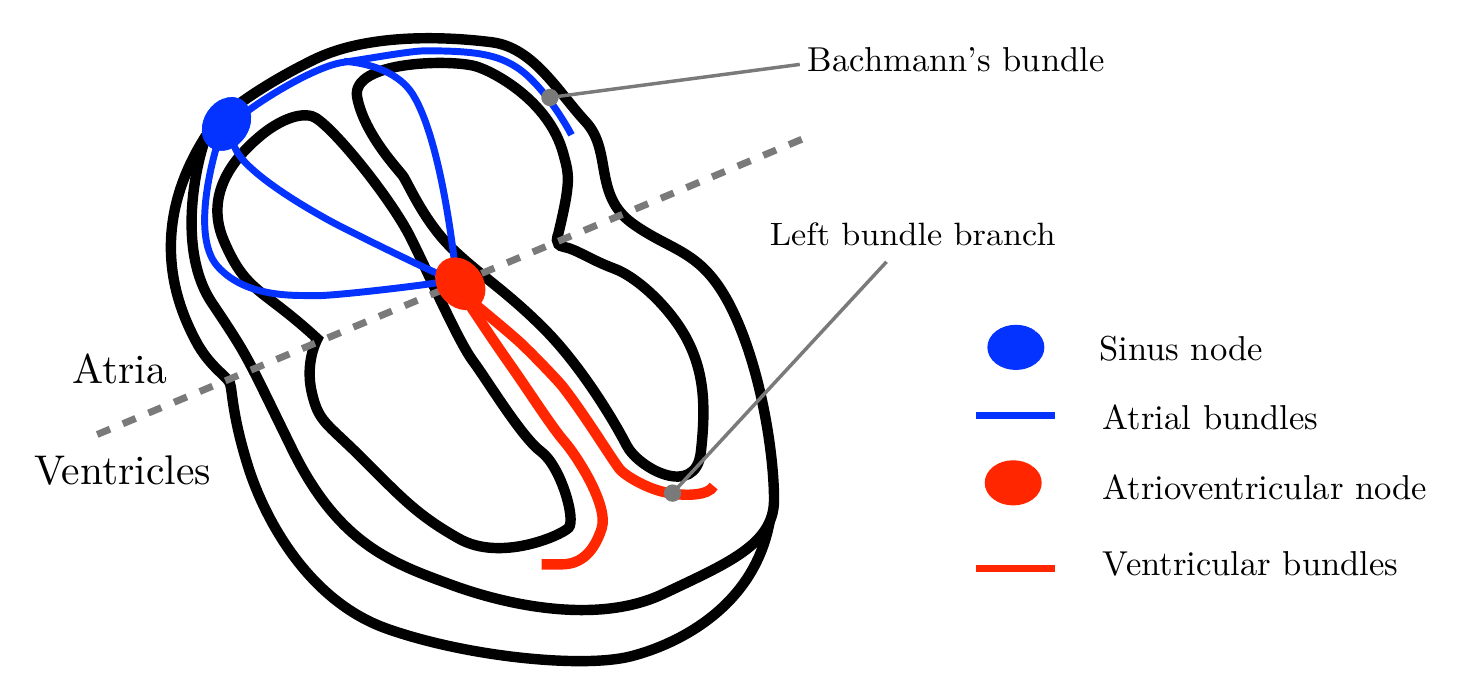}
	\caption{Heart conduction system}
	\label{fig:normalCase}
\end{figure}

Figure~\ref{fig:normalCase} represents a schematic view of the heart conduction system in a healthy heart: the sinus and atrioventricular nodes, the Bachmann bundle and the Purkinje fibers. In this work, the atrio-ventricular node and the Purkinje fibers are not explicitly modeled (see below). 

%=========================================================================
\section{Modeling assumptions}
\label{sec:modeling_assump}
In this section, we present the electrophysiology equations and the ionic models used in the ventricles and the atria. We also present the coupling conditions between the atria and the ventricles and between the heart and the body.

%==============
\subsection{Bidomain model}\label{subsec:bidMod}

In order to describe the electrical potential in the heart, we use the standard nonlinear reaction-diffusion equations known as the bidomain equations \cite{sachse-04,sundnes-lines-cai-06}. In terms of extracellular potential $\ue $ and transmembrane potential $\vm  = \ui  - \ue $, with $\ui $ the intracellular potential, the bidomain model reads
\begin{equation} 
	\label{eq:bidomainmodel-3D}
	\left \{
	\begin{array}{rcl}
		 \dsp A_m\Bigl(C_m \frac{\partial V_m}{\partial t} + I_{ion}(V_m,w_1, \ldots, w_n) \Bigr) - \div\bigl(\vec{\vec{\sigma}}_i  \cdot \vec{\nabla}V_m\bigr)  &=&  \div\bigl(\vec{\vec{\sigma}}_i \cdot \vec{\nabla}u_e\bigr) + A_m I_{app}, \\
	 	\dsp \div\Bigl(\bigl(\vec{\vec{\sigma}}_i + \vec{\vec{\sigma}}_e\bigr) \cdot \vec{\nabla}u_e\Bigr) 
			&=&  - \div\bigl( \vec{\vec{\sigma}}_i \cdot \vec{\nabla}V_m\bigr), \\
	\end{array}
	\right.
\end{equation}
in $ \B \times (0,T)$, where $\B$ denotes the 3D domain of interest, $A_m$ is a positive constant denoting the ratio of membrane area per unit volume, $C_m$ the membrane capacitance per unit surface, $I_{ion}$ the ionic current which depends on $n$ ionic variables $w_1$, \ldots, $w_n$ and $I_{app}$ a given applied stimulus current.

We assume that the heart is isolated, so we make the standard assumption that the extracellular current does not flow through the epicardium:
\begin{equation} \label{eq:bidomainmodelbc1}
	\begin{array}{rcll}
		\bigl(\vec{\vec{\sigma}}_e \cdot \vec{\nabla} \ue  \bigr) \cdot \vec{n} & =& 0, & \mbox{ in } \partial \B \times (0,T).	
	\end{array}
\end{equation}
The second boundary condition comes from the fact that, by definition, the intra-cellular current does not propagate outside the heart~\cite{tung-78}:
\begin{equation} \label{eq:bidomainmodelbc2}
	\begin{array}{rcll}
		\bigl(\vec{\vec{\sigma}}_i\cdot \vec{\nabla} \ue  \bigr) \cdot \vec{n} & =& -\bigl(\vec{\vec{\sigma}}_i\cdot \vec{\nabla} \vm  \bigr) \cdot \vec{n}, & \mbox{ in }  \partial \B \times (0,T).
	\end{array}
\end{equation}

In order to define the $I_{ion}$ term, equations \eqref{eq:bidomainmodel-3D} must be coupled with a ionic model, i.e. a system of nonlinear ordinary differential equations (ODEs). For the ventricular domain, we use the MV model, which is a phenomenological model associated with three ionic currents, three gate variables, and governed by 28 parameters~\cite{buenoOrovio-cherry-fenton-08}.

In order to include the anisotropy between the orthogonal and the tangent directions of the fibers, the conductivity tensors $\TensorThreeD_i$ and $\TensorThreeD_e$ are defined by
\[
	\TensorThreeD_{i,e} = \sigma_{i,e}^{v,t} \, \vec{\vec{I}}  + (\sigma_{i,e}^{v,l} -\sigma_{i,e}^{v,t} ) \, \vec{\tau} \otimes \vec{\tau},
\]
where $\vec{\vec{I}}$ denotes the 3D identity matrix, the vector $\vec{\tau}$ is of unit length and parallel to the local fiber direction, and $\sigma_{i,e}^{v,l} $ and $\sigma_{i,e}^{v,t} $ are respectively the conductivity coefficients in the intra- and extra-cellular ventricular medium measured along and across the fiber direction.

The bidomain model can be rewritten in weak form as follows. For all $t>0$, find $V_m(\cdot,t) \in H^1(\B)$, $u_e(\cdot,t) \in H^1(\B)$ and $w_1(\cdot,t), \ldots, w_n(\cdot,t) \in L^{\infty}(\B)$ with $\int_{\B} u_e = 0$, such that
\begin{equation} \label{eq:bidomainmodel-3D-var}
	\left \{
	\begin{array}{l}
		A_m  \dsp \int_{\B} \Bigl(  C_m  \frac{\partial V_m}{\partial t}  +  I_{ion}(V_m,w_1,\ldots,w_n) \Bigr)  \phi   
		+   \int_{\B} \Bigl[ \TensorThreeD_i \cdot   \bigl(\vec{\nabla}V_m + \vec{\nabla}u_e\bigr) \Bigr] \cdot  \vec{\nabla}\phi  \\
		\dsp \hspace{7cm} =   A_m \dsp\int_{\B} I_{app} \phi, \\
  		\dsp \int_{\B} \Bigl[ (\TensorThreeD_i + \TensorThreeD_e) \cdot \vec{\nabla}\ue   \Bigr] \cdot  \vec{\nabla} \psi 
			+  \int_{\B} \Bigl[ \TensorThreeD_i \cdot \vec{\nabla}\vm  \Bigr] \cdot  \vec{\nabla} \psi =  0,\\
	\end{array}
	\right.
\end{equation}
for all $\phi, \psi \in H^1(\B)$ such that $\int_{\B} \psi = 0$. Under some regularity assumptions, we have existence and uniqueness of a solution of the bidomain model~\cite{boulakia-fernandez-gerbeau-zemzemi-08,bourgault-coudiere-pierre-09}. The hypothesis $\int_{\B} u_e = 0$ is necessary in order to have uniqueness and we show in Section~\ref{sec:wholeHeart} how to adapt this condition when atria and ventricles are coupled. Note that we do not model the blood within the atria and ventricles.

\subsection{Surface bidomain model}\label{subsec:surfBidMod}
As explained in Section~\ref{sec:mesh}, it is more convenient to work with a surface mesh for the atria. The electrophysiology model used in the present work for the atria was derived from the volume bidomain model and set on the midsurface. It was obtained from a rigorous asymptotic analysis \cite{chapelle-collin-gerbeau-12} and was specifically designed for thin cardiac structures \cite{collin-gerbeau-hocini-13}.  Compared to its 3D counterpart, it is very attractive from a computational standpoint. It is worth emphasizing that, even if it is set on the midsurface, it actually takes into account volume effects, like the strong anisotropy variations across the thickness. In \cite{chapelle-collin-gerbeau-12}, we have also performed a numerical assessment of the surface model by comparing the results given by the 3D bidomain model and the surface model for different geometries and we have used realistic values characteristic of atrial electrophysiology for all dimensions and parameters. In particular, we have showed that with a thickness of $0.2$mm --~which can be seen as a standard atrial wall thickness, see \cite{beinart-abbara-blum-11}~-- the agreement between 3D and surface models is excellent.

Let $\Surf$ denote the midsurface of the wall and $H^1(\Surf)$ the associated functional space.
The surface-based bidomain model can be rewritten in weak form as follows: for all $t>0$, find $V_m(\cdot,t) \in H^1(\Surf)$, $u_e(\cdot,t) \in H^1(\Surf)$ and $w_1(\cdot,t), \ldots, w_n(\cdot,t) \in L^{\infty}(\Surf)$ with $\int_{\Surf} u_e = 0$, such that
\begin{equation} \label{eq:bidomainmodel-curv}
	\left \{
	\begin{array}{l}
		A_m  \dsp\int_{\Surf} \Bigl( C_m  \frac{\partial V_m}{\partial t}  +  I_{ion}(V_m,w_1, \ldots, w_n) \Bigr)  \phi    
		+  \dsp\int_{\Surf} \Bigl(\TensorTwoD\mbox{}_{\, i} \cdot   \bigl(\underline{\nabla}V_m 
		+ \underline{\nabla}u_e\bigr)\Bigr)\cdot  \underline{\nabla}\phi   \\
		\hspace{7cm}=  A_m   \dsp\int_{\Surf} I_{app} \phi  , \\
  		\dsp\int_{\Surf} \Bigl((\TensorTwoD\mbox{}_{\, i} + \TensorTwoD\mbox{}_{\, e}) \cdot \underline{\nabla}\ue  \Bigr) \cdot  \underline{\nabla} \psi    
		+  \dsp \int_{\Surf} \Bigl(\TensorTwoD\mbox{}_{\, i} \cdot \underline{\nabla}\vm  \Bigr) \cdot  \underline{\nabla} \psi   =  0, \\
	\end{array}
	\right.
\end{equation}
for all $\phi, \psi \in H^1(\Surf)$ such that $\int_{\Surf} \psi = 0$. We denote by $\sigma_{i,e}^{a,l} $ and $\sigma_{i,e}^{a,t} $ the conductivity coefficients in the intra- and extra-cellular atrial medium measured along and across the fiber direction.
The intra- and extra-cellular diffusion tensors $\TensorTwoD\mbox{}_{\, i}$ and $\TensorTwoD\mbox{}_{\, e}$ are defined by
\begin{equation}\label{eq:sigma_ie}
	\TensorTwoD\mbox{}_{\, i, e} = \sigma_{i,e}^{a,t}  \underline{\underline{I}} +  (\sigma_{i,e}^{a,l}  - \sigma_{i,e}^{a,t} )  \bigl[ I_0(\theta) \underline{\tau}_0 \otimes \underline{\tau}_0 + J_0(\theta) \underline{\tau}_0^\perp \otimes \underline{\tau}_0^\perp \bigr],
\end{equation}
where $\underline{\underline{I}}$ denotes the identity tensor in the tangential plane, $\underline{\tau}_0$ is a unit vector parallel to the local fiber direction on the atria midsurface, and $\underline{\tau}_0^\perp$ such that $(\underline{\tau}_0,\underline{\tau}_0^\perp)$ gives an orthonormal basis of the tangential plane. We use the fibers direction at the endocardium and at the epicardium to define the fibers direction $\underline{\tau}_0$ on the atria midsurface and the angle variation $\theta$ between the endocardium and the epicardium. The effect of angular variations appears in the model with the coefficients $I_0(\theta) = \frac{1}{2} + \frac{1}{4 \theta} \sin(2 \theta)$ and $J_0(\theta) = 1 - I_0(\theta)$. Note that $J_0(\theta) = 0$  (and $I_0(\theta) = 1$) if and only if $\theta=0$, which corresponds to a constant direction in the thickness and then $\TensorTwoD\mbox{}_{\, i, e} = \sigma_{i,e}^{a,t}  \underline{\underline{I}} +  (\sigma_{i,e}^{a,l}  - \sigma_{i,e}^{a,t} ) \, \underline{\tau}_0 \otimes \underline{\tau}_0$. By contrast, important angular variations make $I_0$ decrease and $J_0$ increase in \eqref{eq:sigma_ie} and the diffusion becomes  more isotropic.  This model has been compared \cite{collin-gerbeau-hocini-13} to several 3D models proposed in the literature \cite{deng-gong-shou-12,harrild-graig-00,matsuo-lellouche-wright-09}.

The Courtemanche-Ramirez-Nattel ionic model \cite{courtemanche-ramirez-nattel-98} is considered. It includes 12 ionic currents and 20 other variables. The two atria are connected only by two regions, the Bachmann bundle and the \emph{Fossa Ovalis}. We refer to~\cite{collin-gerbeau-hocini-13} for more details.

\subsection{Simulations on the whole heart} \label{sec:wholeHeart}

\paragraph{Coupled model}
%From a mathematical point of view, volume and surface models are incompatible. It would be erroneous to solve them separately because the uniqueness criterion for the first model is not consistent with the second one. 
As seen in Sections \ref{subsec:bidMod} and \ref{subsec:surfBidMod}, the unique solution $u_e(\cdot,t) \in H^1(\B)$ of \eqref{eq:bidomainmodel-3D-var} is s.t. $\int_{\B} u_e = 0$ and the unique solution $u_e(\cdot,t) \in H^1(\Surf)$ of \eqref{eq:bidomainmodel-curv} is s.t. $\int_{\Surf} u_e = 0$. Here we consider the whole domain $\B \cup \Surf$ and a new global criterion. The resulting coupled problem is well-posed at the discrete level, but its mathematical analysis remains to be done. Let $\Omega_h = \B_h \cup \Surf_h$, where $\B_h$ is the mesh of the ventricles  and $\Surf_h$ is the mesh of the atria, and let $\mathcal{L}_h$ be the line such that $\B_h \cap \Surf_h = \mathcal{L}_h$. We denote by $\gamma_{\tilde{\Omega}} u$ the restriction of a function $u$ to a subdomain $\tilde{\Omega}$. The  finite dimensional approximation space $\V_h$ is then defined by:  $u_h \in \V_h$ if and only if $u_h$ is continuous in $\Omega_h$, $\gamma_{\B_h} u_h \in H^1(\B_h)$, $\gamma_{\Surf_h} u_h  \in H^1(\Surf_h)$,  and $\int_{\B_h} u_h = 0$. Using \eqref{eq:bidomainmodel-3D-var}, \eqref{eq:bidomainmodel-curv}, the full model reads, find $(u_{e,h}, \, V_{m,h}) \in \V_h$ such that $\forall \phi, \psi \in \V_h$,
\begin{equation} \label{eq:bidomainmodel}
	\left \{
	\begin{array}{l}
		A_m  \dsp\int_{\Surf_h} \Bigl( C_m  \frac{\partial V_{m,h}}{\partial t}  +  I_{ion}(V_{m,h}, \cdots) \Bigr)  \phi    
		+ A_m  \dsp \int_{\B_h} \Bigl(  C_m  \frac{\partial V_{m,h}}{\partial t}  +  I_{ion}(V_{m,h},\cdots))  \phi  \\
		+  \dsp\int_{\Surf_h} \Bigl(\TensorTwoD\mbox{}_{\, i} \cdot   \bigl(\underline{\nabla}V_{m,h} 
		+ \underline{\nabla}u_{e,h}\bigr)\Bigr)\cdot  \underline{\nabla}\phi
		+   \int_{\B_h} \Bigl[ \TensorThreeD_i \cdot   \bigl(\vec{\nabla}V_{m,h} + \vec{\nabla}u_{e,h}\bigr) \Bigr] \cdot  \vec{\nabla}\phi \\
		\hspace{5cm}=  A_m   \dsp\int_{\Surf_h} I_{app} \phi  + A_m \dsp\int_{\B_h} I_{app} \phi, \\
  		\dsp\int_{\Surf_h} \Bigl((\TensorTwoD\mbox{}_{\, i} + \TensorTwoD\mbox{}_{\, e}) \cdot \underline{\nabla}u_{e,h} \Bigr) \cdot  \underline{\nabla} \psi    
		+  \dsp \int_{\Surf_h} \Bigl(\TensorTwoD\mbox{}_{\, i} \cdot \underline{\nabla}V_{m,h} \Bigr) \cdot  \underline{\nabla} \psi   \\
  		\hspace{2cm}+ \dsp \int_{\B_h} \Bigl[ (\TensorThreeD_i + \TensorThreeD_e) \cdot \vec{\nabla}u_{e,h}  \Bigr] \cdot  \vec{\nabla} \psi
			+  \int_{\B_h} \Bigl[ \TensorThreeD_i \cdot \vec{\nabla}V_{m,h} \Bigr] \cdot  \vec{\nabla} \psi =  0.
	\end{array}
	\right.
\end{equation}

\begin{figure}
	\centering
	\includegraphics[width=0.7\textwidth]{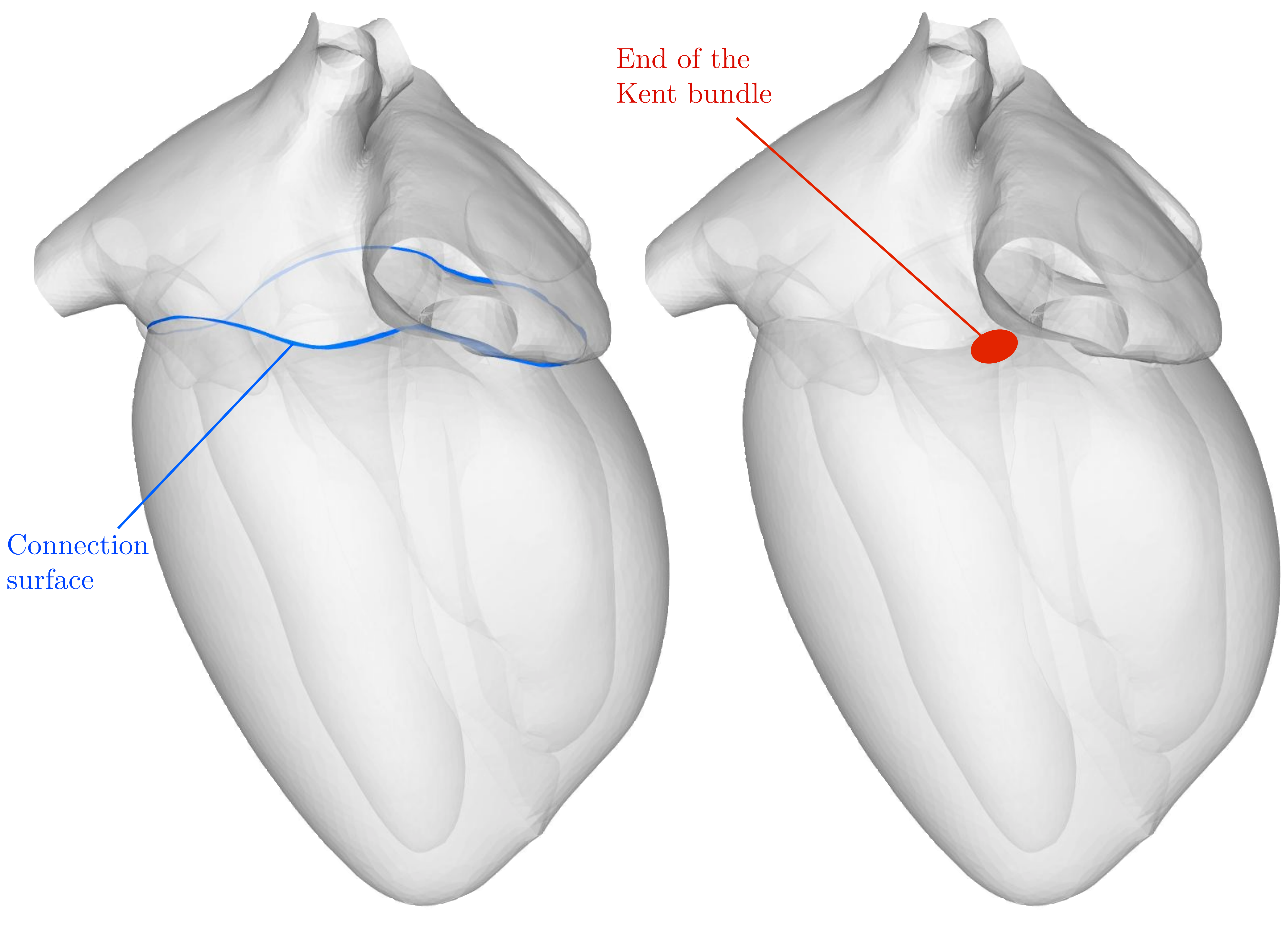}
	\caption{Left - Surface connection between the atria and the ventricles, Right - Kent bundle}
	\label{fig:connectionSurfaceKB}
\end{figure}

\paragraph{Connection surface}
As previously mentioned, the atrioventricular node is the only pathway for the electrical signal between the atria and the ventricles. From a physiological point of view, a fibrous skeleton separates atria boundaries from ventricles epicardium. This layer isolates the atrial cells from the ventricular ones~\cite{martini-timmons-tallitsch-11,mcguire-bakker-vermeulen-96}.  We propose to model this fibrous skeleton with a thin layer of the atrial surface, represented on the left of Figure~\ref{fig:connectionSurfaceKB}. The idea is that in this area there is only a low conduction of the extracellular potential. In this region, denoted by $\Surf_c \subset \Surf$, the intracellular conductivity is set to  zero and the extracellular conductivity denoted by $\sigma_{e}^{s,t}$ is very low (see Table~\ref{tab:parametersTensor}). Finally, the surface and volume bidomain equations~ \eqref{eq:bidomainmodel} are solved simultaneously on this ``hybrid'' domain 
with
\begin{eqnarray*}
	\TensorThreeD_{i,e} = \sigma_{i,e}^{v,t} \, \vec{\vec{I}}  + (\sigma_{i,e}^{v,l} -\sigma_{i,e}^{v,t} ) \, \vec{\tau} \otimes \vec{\tau}, \\
	\TensorTwoD\mbox{}_{\, i, e}  = \sigma_{i,e}^{a,t}  \underline{\underline{I}} +  (\sigma_{i,e}^{a,l}  - \sigma_{i,e}^{a,t} )  \bigl[ I_0(\theta) \underline{\tau}_0 \otimes \underline{\tau}_0 + J_0(\theta) \underline{\tau}_0^\perp \otimes \underline{\tau}_0^\perp \bigr], \mbox{ in } \Surf \setminus \overline{\Surf}_c, \\
	\TensorTwoD\mbox{}_{\, i}  =  \underline{\underline{0}}, \mbox{ and } \TensorTwoD\mbox{}_{\, e} = \sigma_{e}^{s,t}  \underline{\underline{I}}, \mbox{ in } \Surf_c.
\end{eqnarray*}

\begin{table}
\centering
\begin{tabular}{ccc}
\hline
\hline
$\sigma_T^\text{body}$ & $\sigma_T^\text{bones}$ & $\sigma_T^\text{lungs}$\\ 
\hline
\hline
$3.0\, 10^{-4}$ & $1.2\, 10^{-4}$ & $2.0\, 10^{-5}$ \\
\hline
\end{tabular}
\caption{Torso conductivity parameters (all in $\textrm{S.cm}^{-1}$)}
\label{tab:bodyPar}
\end{table}

\paragraph{Parameters and applied currents}
The values of the membrane parameters are $A_m = 200.0 \, \textrm{cm}^{-1}$ and $C_m = 10^{-3} \;  \textrm{mF.cm}^{-2}$ for the whole heart. The conductivity takes different values depending on the region in the ventricles and atria (Table~\ref{tab:parametersTensor}).
\begin{table}
\centering
\begin{tabular}{ccccccccc}
\hline
\hline
$\sigma_e^{v,t}$ & $\sigma_e^{v,l}$ & $\sigma_i^{v,t}$ & $\sigma_i^{v,l}$ & $\sigma_e^{a,t}$ & $\sigma_e^{a,l}$ & $\sigma_i^{a,t}$ & $\sigma_i^{a,l}$ & $\sigma_e^{s,t}$\\ 
\hline
\hline
%$6.0\, 10^{-4}$ & $2.0\, 10^{-3}$ & $2.0\, 10^{-4}$ & $2.0\, 10^{-3}$ & $9.0\, 10^{-4}$ & $2.5\, 10^{-3}$ & $2.5\, 10^{-4}$ & $2.5\, 10^{-3}$ &  $7.5\, 10^{-7}$ \\
$4.8\, 10^{-4}$ & $1.6\, 10^{-3}$ & $1.6\, 10^{-4}$ & $1.6\, 10^{-3}$ & $9.0\, 10^{-4}$ & $2.5\, 10^{-3}$ & $2.5\, 10^{-4}$ & $2.5\, 10^{-3}$ &  $7.5\, 10^{-7}$ \\
\hline
\end{tabular}
\caption{Conductivity parameters (all in $\textrm{S.cm}^{-1}$)}
\label{tab:parametersTensor}
\end{table}

In the atria,  the regions of fast conduction are the Bachmann bundle (BB), see Figure~\ref{fig:normalCase}, the \emph{Crista Terminalis} (CT) and the pectinate muscles (PM). By contrast, the \emph{Fossa Ovalis} (FO) is a region of slow conduction. 
In order to model the different propagation velocities, we modify the values of $g_{Na}$, the maximal conductance of the $Na^{2+}$ current $I_{Na}$. Table~\ref{tab:conductance} gives the parameters used for $g_{Na}$. Furthermore, to reduced the action potential duration,  the parameter $g_{k_s}$ is chosen five times as big as in the original model~\cite{courtemanche-ramirez-nattel-98}.

In the ventricles, we modify the duration of the plateau too. In the MV model, we change the values of $\tau_{so_1}$ parameter in order to reduce the action potential duration for epicardial, endocardial and midmyocardial cells. This heterogeneity is considered in the left ventricle, for the positivity of T wave~\cite{yan-antzelevitch-98,keller-weiss-seemann-12}. In the right ventricle, the cells are considered homogeneous and their parameters are taken as in the left ventricle epicardium, except for $\tau_{so_1}$ (Table~\ref{tab:tau-so-1}).

\begin{table}
\centering
\begin{tabular}{ccccccc}
\hline\hline
regular tissue & PM & CT & BB & FO\\ 
\hline
\hline
$7.8$  & $11.7$ & $31.2$ & $46.8$ & $3.9$\\
\hline
\end{tabular}
\caption{Maximal conductance $g_{Na}$ in the different atrial areas (all in $\textrm{nS}.\textrm{pF}^{-1}$)}
\label{tab:conductance}
\end{table}

\begin{table}
\centering
\begin{tabular}{c|cccccc}
\hline\hline
& EPI & ENDO & M & RV\\ 
\hline
\hline
\cite{buenoOrovio-cherry-fenton-08}  & $30.0181$ & $40.0$ & $91.0$ & / \\
\hline
%heart  & $25.0$ & $40.0$ & $61.0$ & $26.0$\\
heart  & $19.0$ & $30.0$ & $45.0$ & $20.0$\\
\hline
\end{tabular}
\caption{Changed ionic parameter $\tau_{so_1}$ of MV compared to~\cite{buenoOrovio-cherry-fenton-08}}
\label{tab:tau-so-1}
\end{table}

\begin{figure}[ht!]
	\centering
	\includegraphics[width=0.85\textwidth]{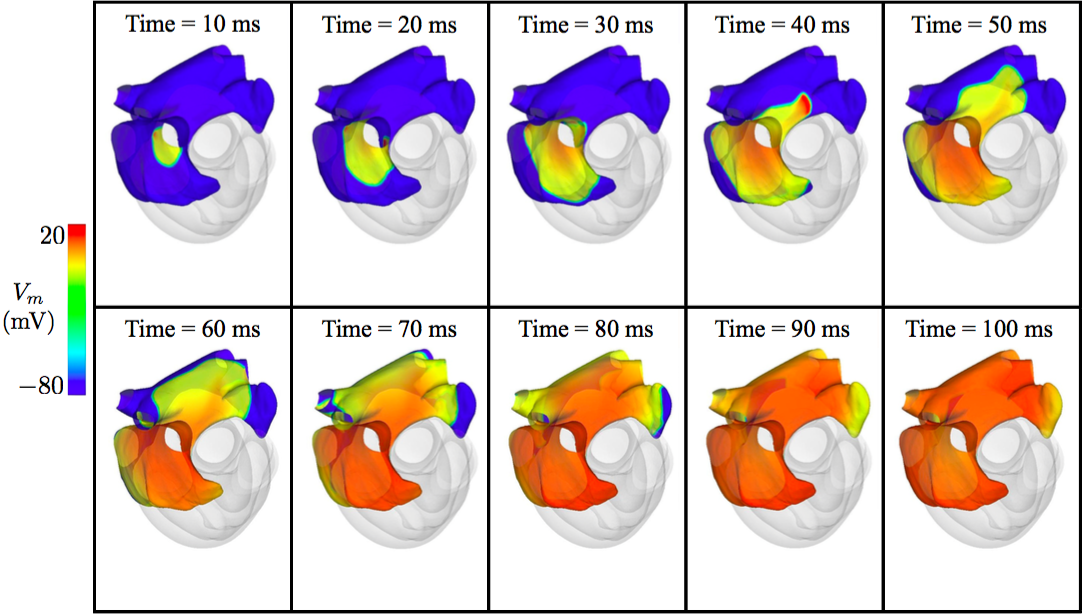}
	\caption{Action potential in the atria.}
	\label{fig:atria}
\end{figure}

\begin{figure}[ht!]
	\centering
	\includegraphics[width=0.9\textwidth]{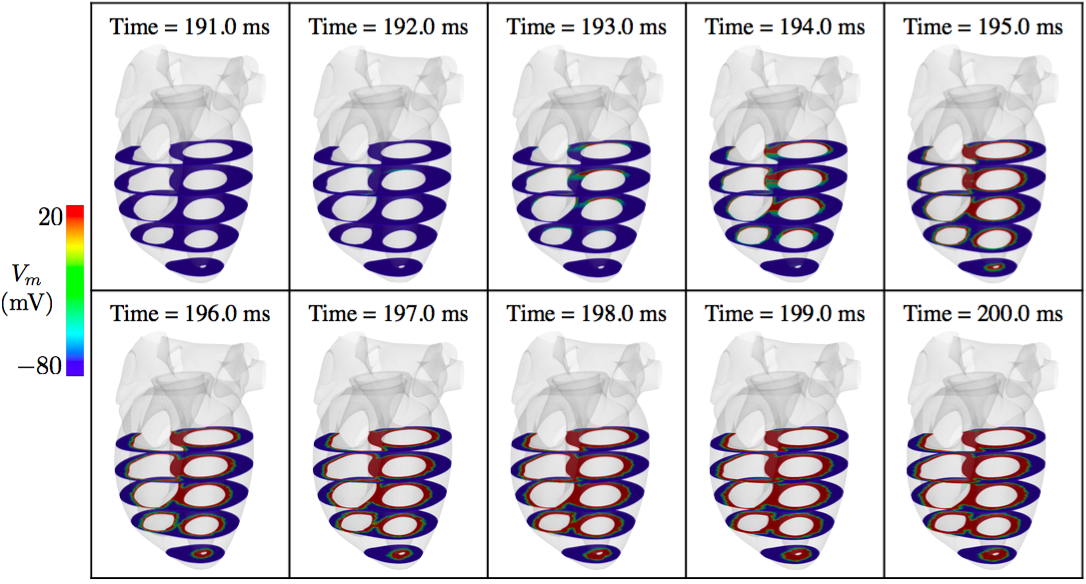}
	\caption{Action potential in the ventricles. A stimulus is applied from t=190 ms to t=195 ms.}
	\label{fig:ventricles}
\end{figure}

Activation is initiated at the sinus node with a stimulus of $2\textrm{ms}$ which triggers a depolarization wavefront in the atria  (Figure~\ref{fig:normalCase}). For the sake of simplicity, the atrioventricular node, which is the only electrical connection between the atria and the ventricles, is not modeled with a sophisticated physiological model. Instead, the excitation is triggered in the ventricle after a parameterized delay (in healthy condition, we choose to start it at 190 ms). Similarly, the fast conduction in the Purkinje fibers (Figure~\ref{fig:normalCase}) is modeled with a predefined stimulus pattern: a time-dependent thin subendocardial layer is activated by an external current for 5 ms on both right and left ventricles~\cite{boulakia-cazeau-fernandez-10}. Figure~\ref{fig:atria} shows the propagation of the activation in the atria and Figure~\ref{fig:ventricles} shows the early stage of the activation in the ventricles.

\begin{figure}
 	\centering
 	\includegraphics[width=\textwidth]{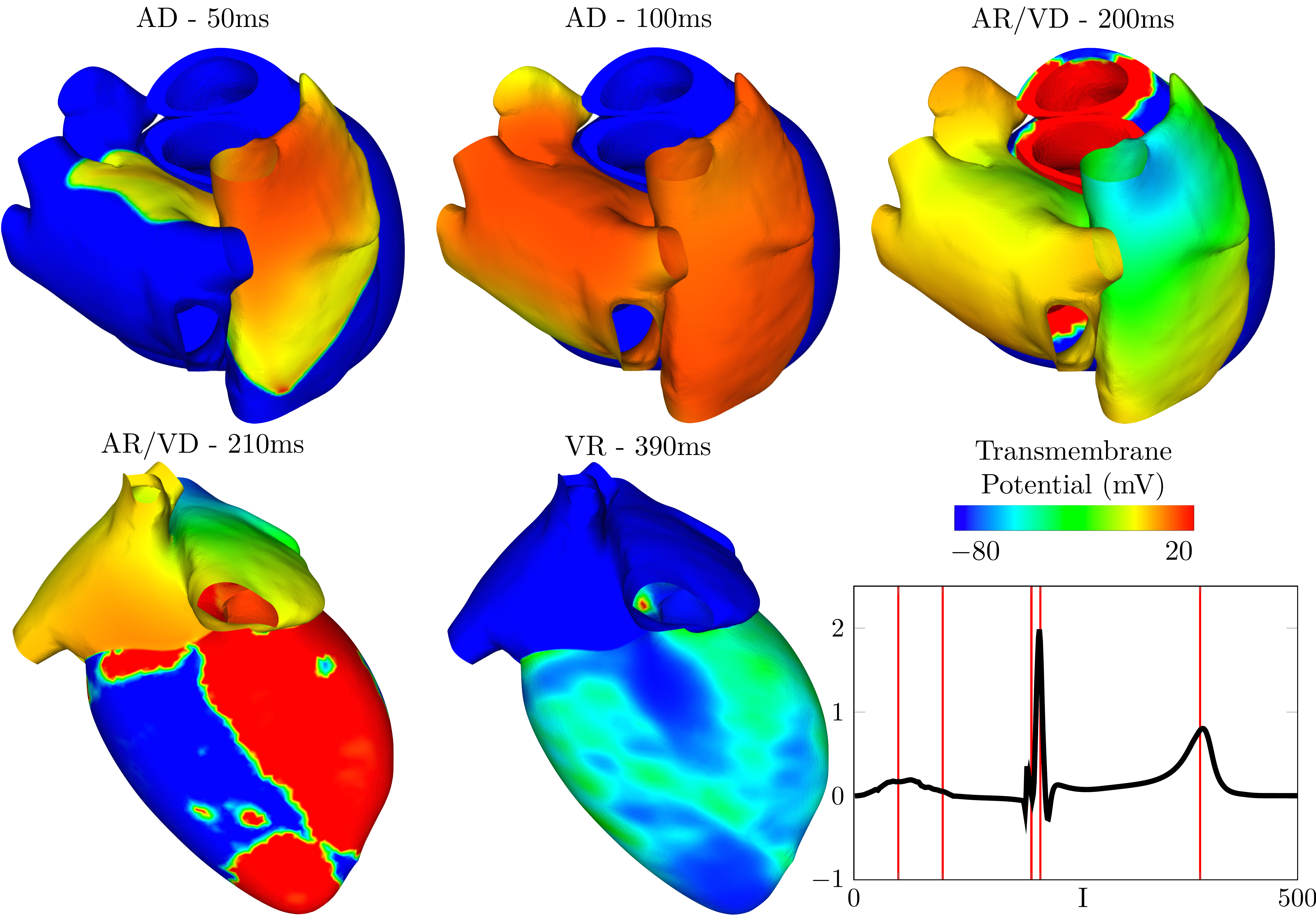}
 	\caption{Simulations of heart depolarization in a healthy case with the corresponding electrocardiogram first lead}
 	\label{fig:simuCompHeart}
 \end{figure}

\paragraph{Simulation results}
The various simulations of this article are performed with the finite element library \emph{FELiScE}\footnote{\url{http://felisce.gforge.inria.fr}}, developed at Inria.  Problem~\eqref{eq:bidomainmodel} is solved with a BDF2 semi-implicit scheme~\cite{boulakia-cazeau-fernandez-10}. Figure~\ref{fig:simuCompHeart} shows a full cardiac cycle. The corresponding first lead electrocardiogram is also represented. The electrical signal starts at the sinus node where the atrial depolarization (AD) begins. By $50\,$ms the wave finishes spreading along the \emph{Crista Terminalis} as a consequence of the high conductivity in this part, see Figure~\ref{fig:simuCompHeart}. Importantly, because of the rapid conduction in the Bachmann bundle, the wave spreads to the left atrial appendage and activates a substantial part of the left atrial wall. The depolarization of the right and left atria terminates at $100\,$ms and $110\,$ms, respectively. The ventricular depolarization begins at $190\,$ms. During this period, the atrial repolarization (AR) occurs. As we can see in Figure~\ref{fig:simuCompHeart}, at $200\,$ms the endocardium of the ventricles rapidly depolarizes. Then, the wave propagates across  the ventricles.  The repolarization ends at $400\,$ms in the right ventricle and at $430\,$ms in the left ventricle.

 \begin{figure}
 	\centering
 	\includegraphics[width=0.75\textwidth]{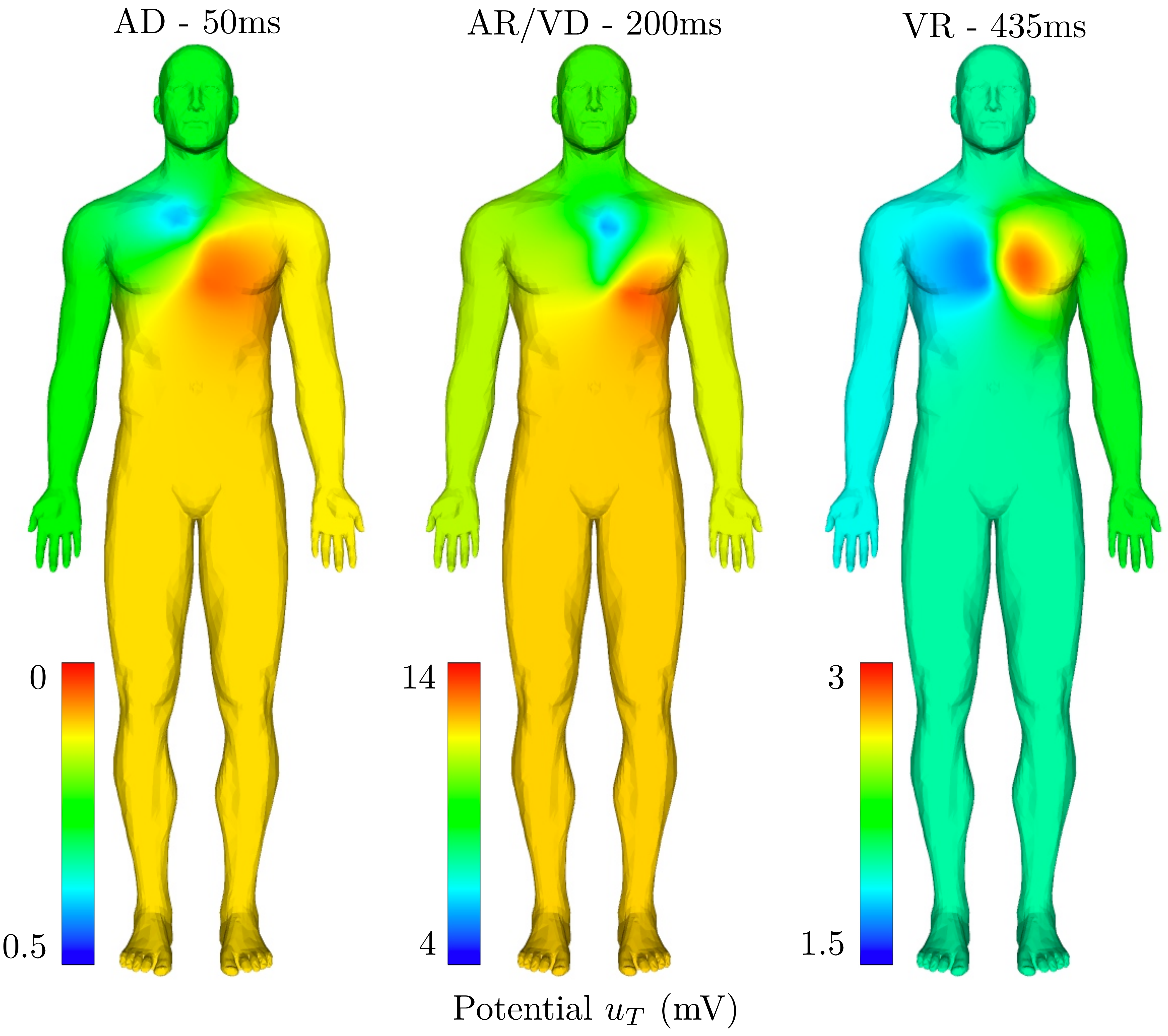}
 	\caption{Body potential simulations of heart depolarization in a healthy case (Figure~\ref{fig:simuCompHeart}), with coupling condition~\eqref{eq:resistcapaccouplWeak}}
 	\label{fig:simuCompBody}
 \end{figure}

\subsection{Coupling with the body}\label{subsec:bodyCR}

The last step in order to obtain an electrocardiogram is to couple the heart model with a diffusion problem in the rest of the body
\begin{equation}\label{eq:bodyDiff}
	-\div (\sigma_T \vec{\nabla}u_T) = 0, \mbox{ in } \Omega_T,
\end{equation}
where the electrical conductivity $\sigma_T$ takes different scalar values in the ribs and the lungs (see \cite{buist-pullan-03} and Table~\ref{tab:bodyPar}). 

On the body surface $\partial \Omega_T^{ext}$, an homogeneous Neumann boundary condition is imposed $\sigma_T \vec{\nabla}u_T \cdot \vec{n} = 0$. To define the transmission conditions at the heart-body interface $\partial \Omega_H$,  we assume that the extracellular current does not flow through the pericardium (isolated heart assumption): 
\begin{equation}\label{eq:isolated}
\vec{\vec{\sigma}}_e \cdot \vec{\nabla} \ue  \cdot \vec{n}=0
\end{equation}
and we consider the resistor-capacitor conditions \cite{boulakia-cazeau-fernandez-10}:
\begin{equation}\label{eq:resistcapaccouplWeak}
	\displaystyle R_p (\sigma_T \vec{\nabla} u_T) \cdot \vec{n} = R_p C_p \frac{\partial(\ue  - u_T)}{\partial t} + (\ue  - u_T)
\end{equation}
where $C_p$ and $R_p$ stand for the capacitance and resistance of the pericardium, respectively. 

Condition \eqref{eq:isolated} is an approximation that is \modify{known}{supposed} not to affect too much the shape of the ECG, as shown in~\cite{boulakia-cazeau-fernandez-10}. It allows us to solve the heart-body system as a one-way coupled problem, which dramatically reduces its computational cost.

%\begin{equation}\label{eq:resistcapaccoupl}
%	\left\{
%	\begin{array}{l}
%	\displaystyle R_p (\sigma_T \vec{\nabla} u_T) \cdot \vec{n} = R_p C_p \frac{\partial(\ue  - u_T)}{\partial t} + (\ue  - u_T), \\
%	(\vec{\vec{\sigma}}_e \cdot \vec{\nabla} \ue  ) \cdot \vec{n} = (\sigma_T \cdot \vec{\nabla} u_T ) \cdot \vec{n},
%	\end{array}
%	\right.
%\end{equation}

Condition \eqref{eq:resistcapaccouplWeak} is a generalization of the condition that is usually adopted (continuity of the potential). It allows us to model the fact that the transmission of potential through the pericardium is not perfect, and can be different for the ventricles and the atria. We take $R_p=10^2~\Omega.\textrm{cm}^{2}$ on the surface in contact with the ventricles and $R_p = 10^5~\Omega.\textrm{cm}^{2}$ on the surface in contact with the atria. These values are chosen to obtain correct relative amplitudes of the P and R waves. We neglect the capacitor effect by taking $C_p = 0~\textrm{mF}.\textrm{cm}^{2}$ in \eqref{eq:resistcapaccouplWeak}. The transmission between the heart and the body is therefore modeled as a Robin boundary condition:
\begin{equation}\label{eq:robin-bc}
		\displaystyle R_p (\sigma_T \vec{\nabla} u_T) \cdot \vec{n} + u_T = u_e, \, \partial \Omega_H.
\end{equation}
Conditions \eqref{eq:isolated} and  \eqref{eq:resistcapaccouplWeak} are further commented in Section~\ref{sec:limitations}.
Figure~\ref{fig:simuCompBody} shows the body surface potential corresponding to the simulation shown in Figure~\ref{fig:simuCompHeart}.
 %In next sections we show some applications of these results. In Section~\ref{sec:ecg} the standard 12-leads ECG is computed in order to study healthy and pathological situations involving atria, ventricles and their interactions. Also, a comparison between the combined Courtemanche/MV ionic models and a simpler phenomenological model such as Mitchell and Schaeffer~\cite{mitchell-schaeffer-03} is presented. Finally, in Section~\ref{sec:grid} we simulate a "virtual" electrodes vest and analyze the signal on the body surface.

\subsection{Electrocardiogram computation}\label{subsec:elec-comp}

\begin{figure}
	\centering
	\includegraphics[width=0.6\textwidth]{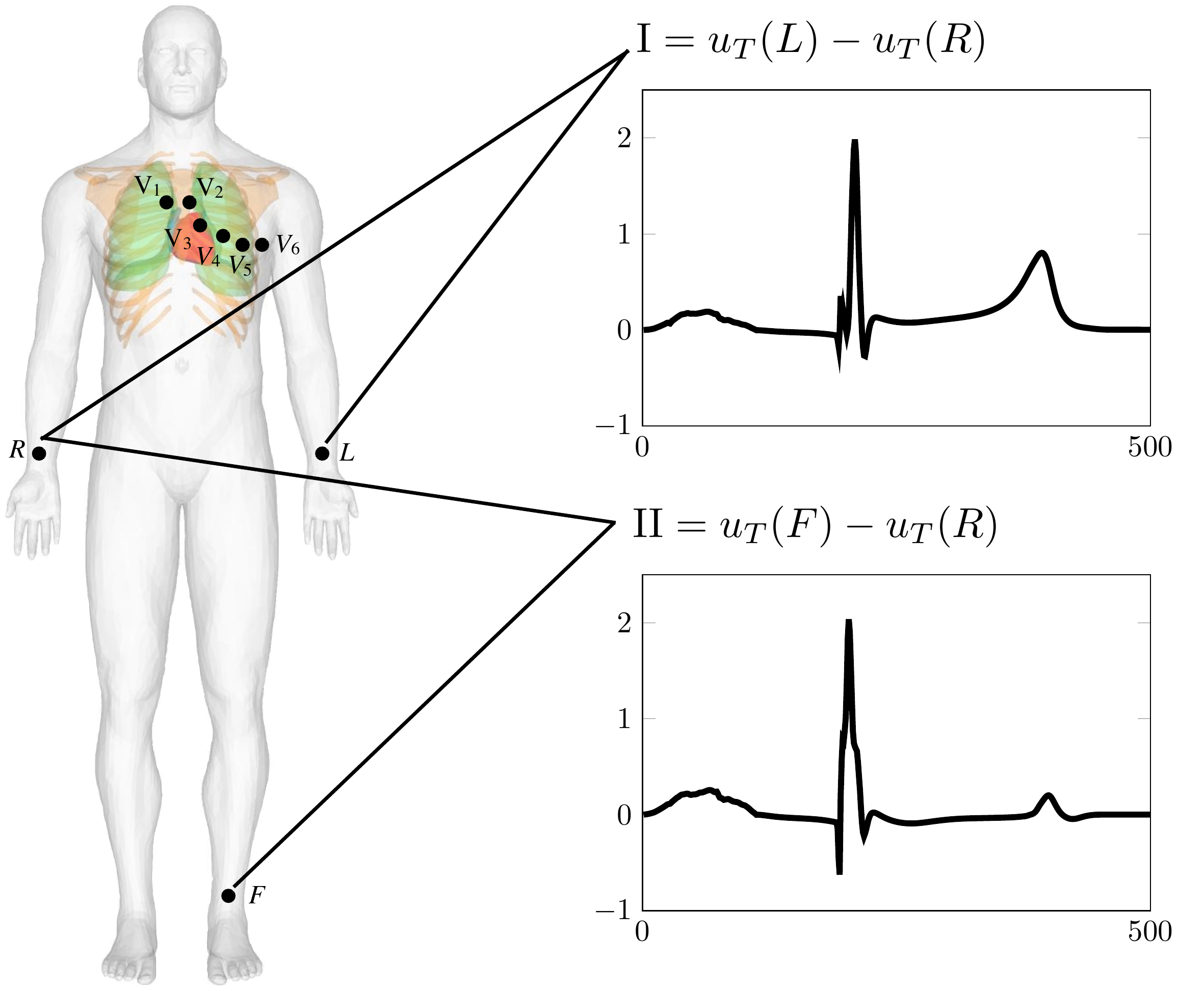}
	\caption{Standard 9 electrodes locations and first and second ECG leads.}
	\label{fig:ecg-elec}
\end{figure}

A standard electrocardiogram is based on the body surface potential recorded by 9 electrodes ($\Gamma_{ECG}=\{R,L,F,V_1,\ldots,V_6\}$, see Figure~\ref{fig:ecg-elec}). These measures are combined to define 12 differences of potential, known as the 12 leads of the standard ECG:
\[
 \begin{array}{l l}
 \text{I}= u_T(L)-u_T(R) & aVR=1.5(u_T(R)-u_w) \\ 
 \text{II}= u_T(F)-u_T(R) & aVL=1.5(u_T(L)-u_w) \\ 
 \text{III}= u_T(F)-u_T(L) & aVF=1.5(u_T(F)-u_w) \\
 V1=u_T(V_1)-u_w & V4=u_T(V_4)-u_w \\
 V2=u_T(V_2)-u_w & V5=u_T(V_5)-u_w \\
 V3=u_T(V_3)-u_w & V6=u_T(V_6)-u_w
 \end{array}
\]
where $u_w=\frac{1}{3}(u_T(L)+u_T(R)+u_T(F))$ is the Wilson potential \cite{malmivuo-plonsey-95}.

%In order to compute the ECG, we are interested in measuring the electrical potential on the body surface. Then the solution of diffusion equations
%\begin{equation}\nonumber
%	-\div (\sigma_T \vec{\nabla}u_T) = 0
%\end{equation}
%in the body domain $\Omega_T$ is needed. As seen in Section~\ref{subsec:bodyCR} different choices are possible for the boundary conditions. In this section we consider a weak coupling with the heart derived from the resistor-capacitor model described above. In particular, boundary conditions~\eqref{eq:resistcapaccouplWeak} are applied with $C_p=0$
%\begin{equation}\label{eq:robin-bc}
%	\left\{
%	\begin{array}{l l}
%		\displaystyle R_p (\sigma_T \vec{\nabla} u_T) \cdot \vec{n} + u_T = u_e, & \partial \Omega_H, \\
%	\sigma_T \vec{\nabla}u_T \cdot \vec{n} = 0, & \partial \Omega_T^{ext}.
%	\end{array}
%	\right.
%\end{equation}

%=========================================================================
\section{Numerical healthy and pathological ECGs} \label{sec:ecg}

In this Section, we present the ECGs provided by the aforementioned model in healthy and pathological conditions. The healthy ECG is obtained by carefully choosing the parameters of the model in order to match most of the qualitative and quantitative features of a physiological ECG. To obtain the pathological ECGs, the approach is different: starting from the nominal values corresponding to a healthy ECG, we modify the parameters in order to model the {\em physical characteristics} of the pathology. Then we observe the effects of these modifications on the numerical ECG, and we compare its features with the ones described in the literature. It is important to emphasize that, for the pathological cases, the parameters are not intentionally fixed to match a given ECG. Thus, if the ECGs obtained after modeling the diseases match the main features observed on real patients, it gives confidence in the prediction capabilities of the model. 

%==============
\subsection{Healthy electrocardiograms}
\begin{figure}
	\centering
	\includegraphics[trim = 15mm 0mm 5mm 0mm, clip, width=\textwidth]{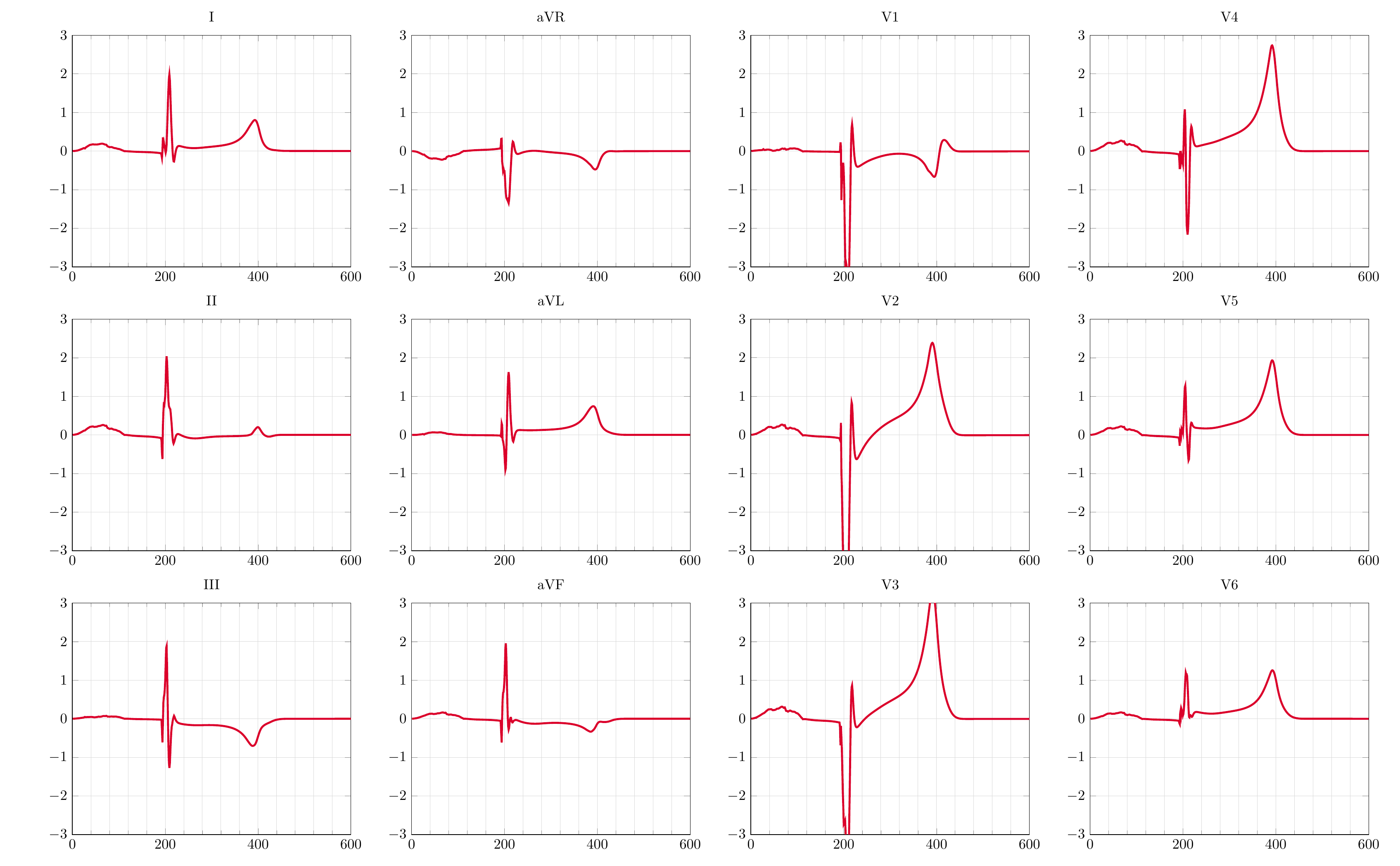}
	\caption{Healthy electrocardiogram corresponding to simulation of Figure~\ref{fig:simuCompHeart} (voltages (mV) versus time (ms))}
	\label{fig:ref}
\end{figure}

Figure~\ref{fig:ref} shows the simulated electrocardiogram in healthy conditions, corresponding to the simulation of Figure~\ref{fig:simuCompHeart}. An electrocardiogram is typically described by distinguishing  five events during the heartbeat, called P, Q, R, S and  T ``waves'' (we will keep this standard terminology even though these events have nothing to do with waves). The P wave corresponds to the atrial depolarization, the QRS complex corresponds to the ventricular depolarization, the T wave corresponds to the ventricular repolarization. The typical durations of each wave, or each interval, are given in Table~\ref{tab:time}. 

Table~\ref{tab:time} also presents the durations of the simulated healthy ECG of Figure~\ref{fig:ref}. These durations are obtained  in the numerical ECG from the landmarks defined according to the following rules: the P wave (resp. QRS complex) starts if $1\%$ of the atria (resp. ventricles) is activated (i.e. if the transmembrane potential $\vm $ is greater than a threshold voltage $V_{gate}$); the P wave (resp. QRS complex) ends when $99\%$ of the atria (resp. ventricles) are activated; the T wave starts when $20\%$ of the ventricles are repolarized (\ie $\vm  \leq 0$); the T wave ends when $99\%$ of the ventricles are fully repolarized ($\vm  \leq 0$). If the minimal value of $\vm $ is $V_{min}=-80\textrm{mV}$ and its maximal value $V_{max} = 20\textrm{mV}$, we define $V_{gate} = -67 \mathrm{mV}$, which corresponds to a threshold voltage $\theta_w = 0.13$ in the MV model. We remark that the waves durations are in general in good accordance with the values found in the literature, even though the QRS complex is slightly shorter than expected.
%We can see in Figure~\ref{fig:ref} that the given values of beginning and end of P and T waves and QRS complex are in good accordance with the ECG.

Table~\ref{tab:criteria} gives the main features of each wave in a normal electrocardiogram. Note that the simulated ECG fulfills almost all the expected criteria.  We only observe a discrepancy in the aVL lead, but this lead is not the most important one for the ECG interpretation.

\begin{figure}
	\centering
	\includegraphics[width=\textwidth]{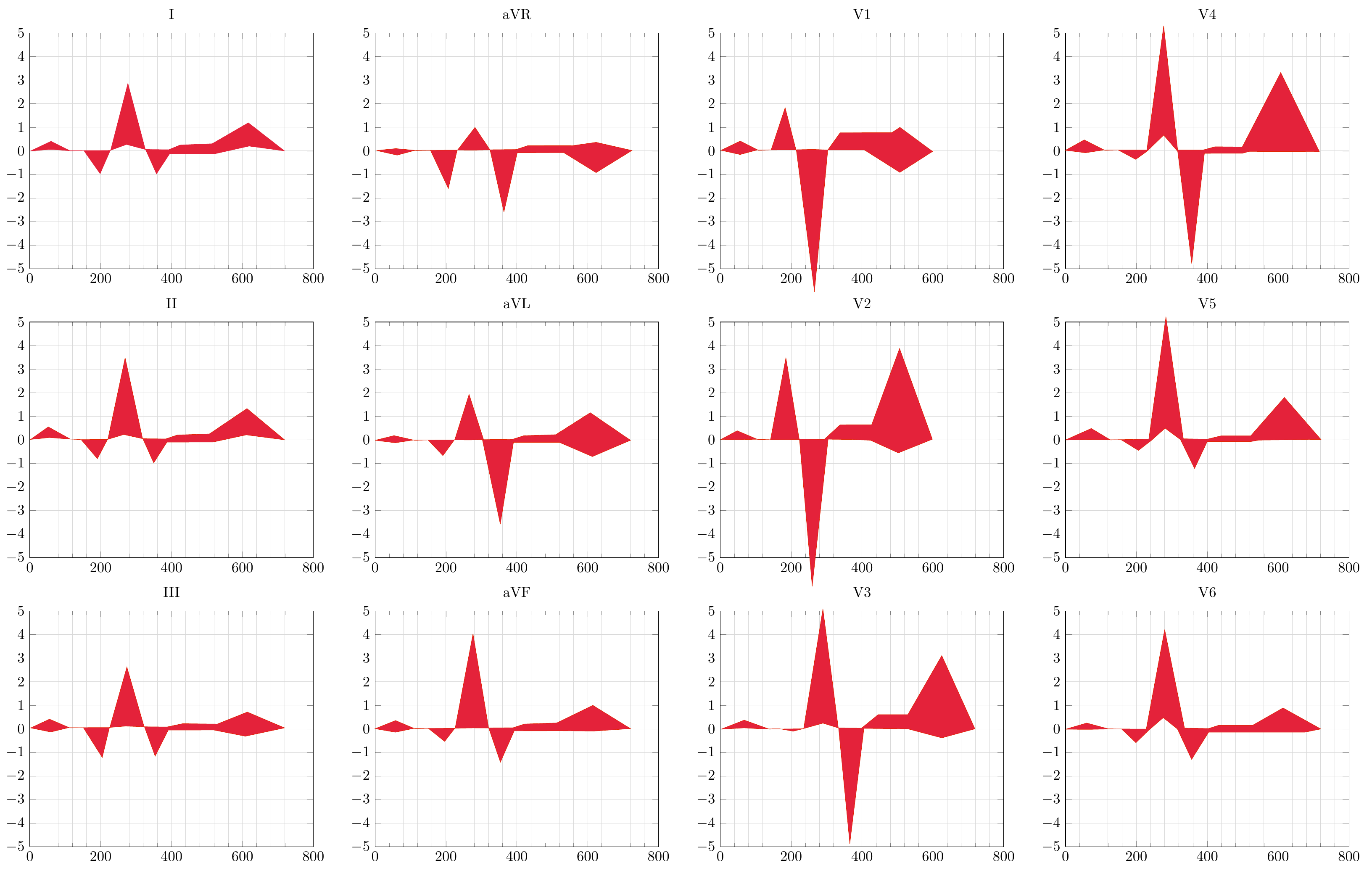}
	\caption{Representation of the variability of the amplitude observed in adults healthy ECGs, according to \cite{wartak-75}.}
	\label{fig:variations}
\end{figure}

To qualitatively assess the waves amplitude and orientation, the schematic presented in Figure~\ref{fig:variations} is very convenient. It is adapted from~\cite{wartak-75} and  shows the normal variations of wave amplitude measured in adults. A visual comparison of Figures~\ref{fig:ref} and~\ref{fig:variations} shows that, for almost every lead, each wave of our numerical ECG is in the range of the normal values. Note that in Figure~\ref{fig:variations}, the length of each wave was arbitrarily chosen as its maximal normal duration. This is the reason why the full PQRST duration is so long in this schematic.

\begin{figure}
	\centering
	\includegraphics[width=0.5\textwidth]{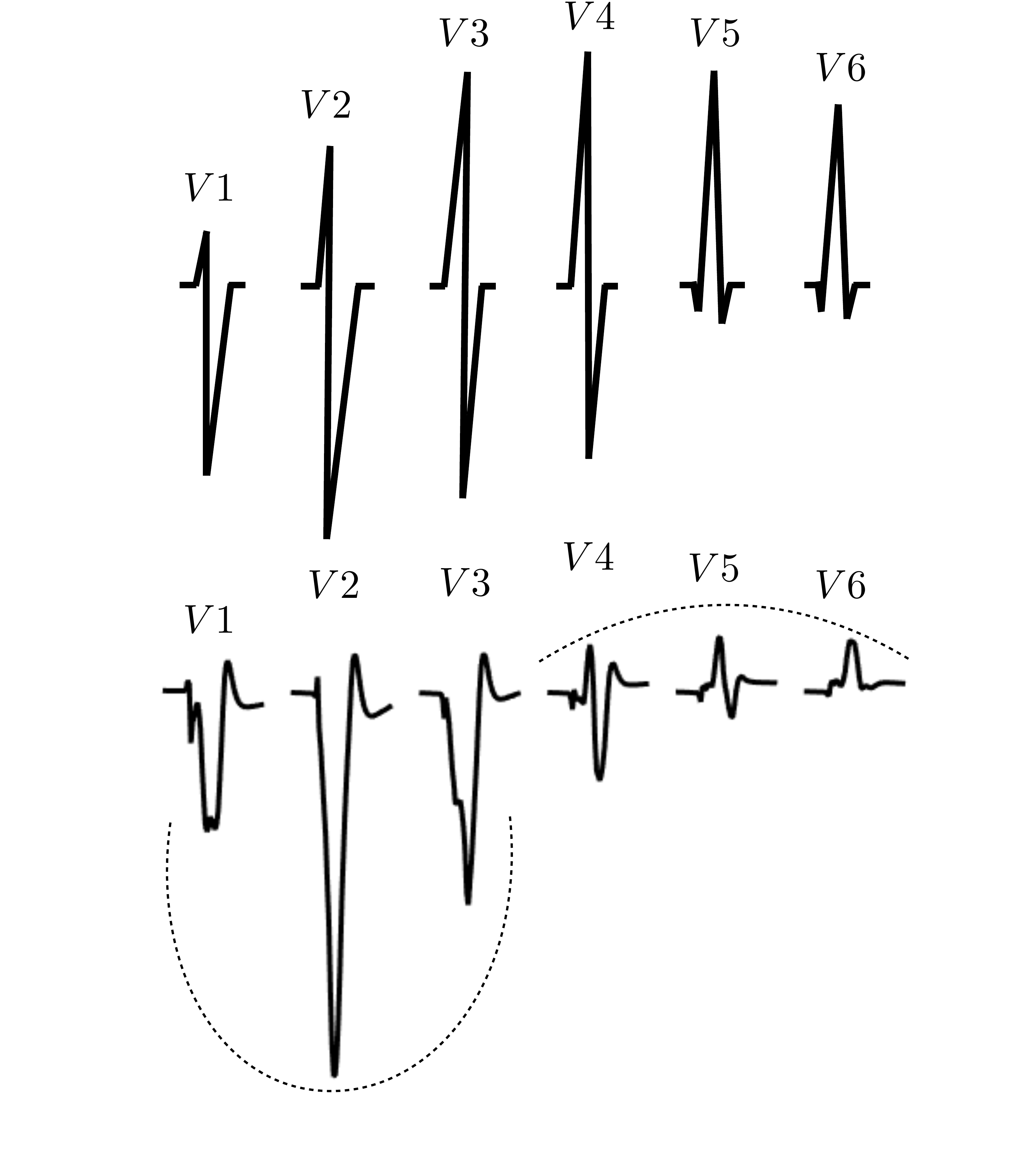}
	\caption{R wave progression in the precordial leads: schematic view from \cite{wartak-75} in the top, and simulated ECG in the bottom}
	\label{fig:Rwave}
\end{figure}

Here is another qualitative assessment. The R wave is known to have an important property in the precordial leads: it uniformly progresses from a rS complex (small R wave, deep S wave) in V1-V2 to a QRS complex in V5-V6 via a RS complex in V3-V4. This represents the rotation  of the cardiac vector from right to left, and from back to front and again back. The idealized behavior is represented in the top of Figure~\ref{fig:Rwave}, extracted from \cite{wartak-75}. The bottom of the same Figure shows the results of our simulation. Given the difficulty to reproduce this subtle dynamics with the simulation of a complex biophysical model, the agreement can be considered as satisfactory.

A last qualitative comment is in order. We note that the P wave presents some oscillations in all the leads of the numerical ECG. The explanation of these oscillations is the brutal changes of the fibers' direction in the atria. It is also possible that the surface representation of the atria accentuates these oscillations by diminishing the natural diffusion arising during the propagation of the depolarization front through the thickness.

\begin{table}
	\centering
	\begin{tabular}{cccccccc}
	\hline\hline
	& P & PR  & Q  & QR  & S  & QRS  & QT \\
	&  wave &  interval &  wave &  interval &  wave &  interval &  interval \\ 
	\hline
	\hline
	Typical & $< 0.12$ & $0.12$  & $<0.04$ & $<0.03$ $V1-V2$ & $<0.04$ & $<0.10$ & $0.35$ \\
 	Duration &  & to $0.21$ && $<0.05$ $V5-V6$ &  & & to $0.45$ \\
	\hline
	\hline
	Healthy & $0.08$ & $0.19$ & $0.015$ & $0.025$ $V1-V2$ & $0.01$ & $0.04$ & $0.25$ \\
	ECG &  &  & & $0.02$ $V5-V6$ & &  & \\
	\hline
	\end{tabular}
	\caption{Durations of the simulated healthy ECG of Figure~\ref{fig:ref} compared with typical durations \cite{wartak-75} (all in s)}
	\label{tab:time}
\end{table}

\begin{table}
	\centering
	\begin{tabular}{ccc}
	\hline\hline
	Wave/Interval & Description & Simulated ECG  \\
	\hline\hline
	 & $\leq 0.25$mV  & $\checkmark$ $0.2$mV  \\
	 P wave &  positive I, II, V3 to V6 & $\checkmark$  \\
	 &  negative aVR &  $\checkmark$  \\
	\hline
	&  limb leads  $\leq 25\%$ of R  & $\checkmark$ \\
	Q wave & precordial leads $\leq 15\%$ of R  & $\checkmark$ \\ 
	&  always negative & $\checkmark$ except for aVL  \\
	\hline
	 & limb leads $\leq 2$mV & $\checkmark$ \\ 
	R wave & precordial leads $\leq 3$mV & $\checkmark$ \\ 
	& always positive, negative in aVR  & $\checkmark$ \\
	& R wave progression, see Figure~\ref{fig:Rwave} & $\checkmark$ \\
	\hline
	& always negative & $\checkmark$ \\ 
	S wave &  small I, II, V5, V6  &  $\checkmark$ \\
	& important V1 to V3 & $\checkmark$  \\
	\hline 
	& $-0.05$mV to $0.1$mV & $\checkmark$ \\
	ST interval & isoelectric  &  $\checkmark$ \\
	& displacement of $0.02$mV in V1, V3 & $\checkmark$  \\
	\hline 
	T wave &  positive I, II, V3 to V6  & $\checkmark$ \\ 
	 &  negative aVR (follow the QRS) &  $\checkmark$  \\
	\hline
	\end{tabular}
	\caption{Criteria for a typical electrocardiogram \cite{wartak-75} compared with simulated ECG of Figure~\ref{fig:ref} }
	\label{tab:criteria}
\end{table}

%==============
\subsection{Pathological electrocardiograms}
\begin{figure}
	\centering
	\includegraphics[width=\textwidth]{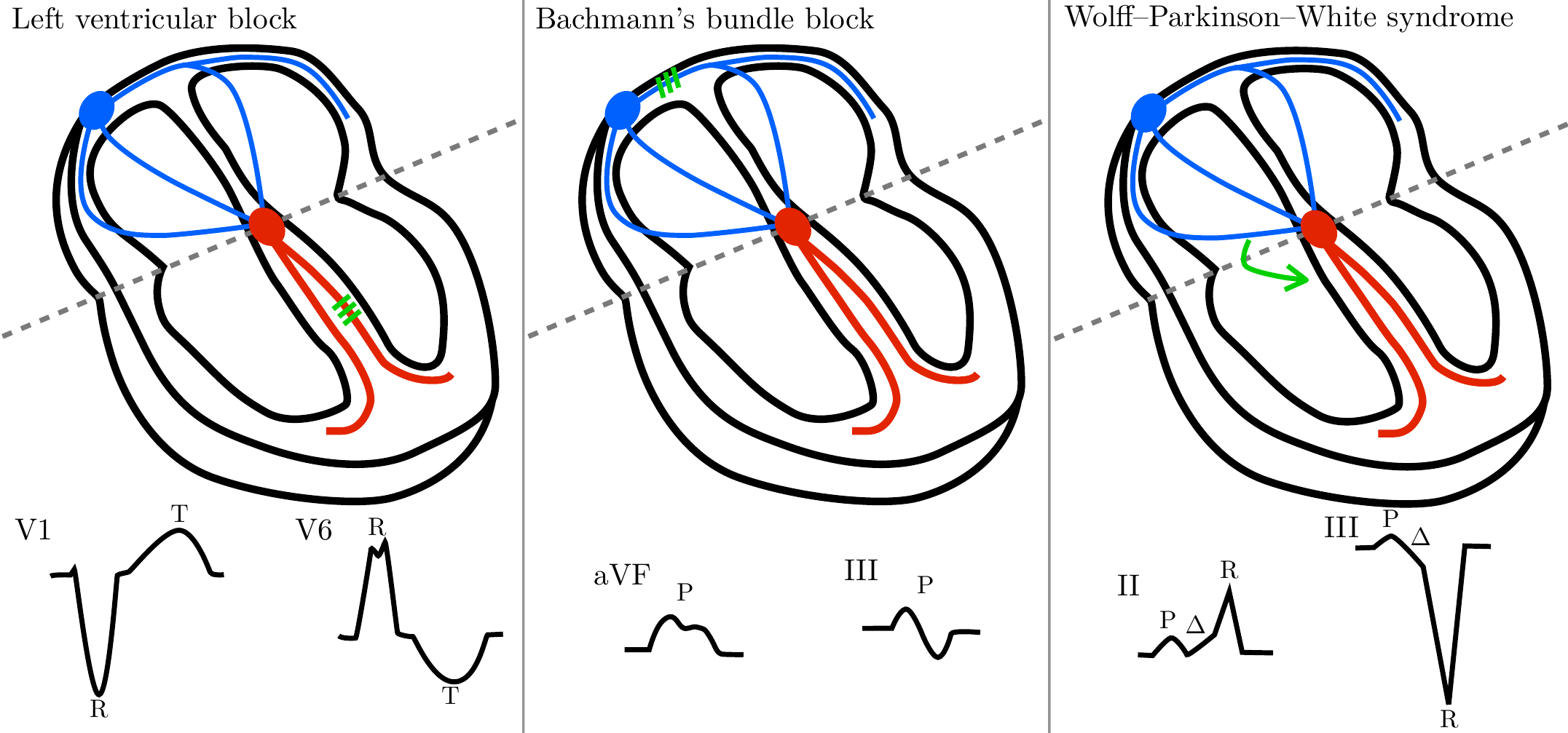}
	\caption{Different pathologies}
	\label{fig:pathologies}
\end{figure}

In this Section, we modify the protocol of the simulation that provided the healthy ECG (Figure ~\ref{fig:ref}) in order to simulate different cardiac pathologies. Then we verify if the numerical ECGs present the main features that allow a medical doctor to detect the pathology. The different pathologies are schematically represented in Figure~\ref{fig:pathologies}, along with the most important leads in each case. 

\subsubsection{Left and right ventricular block}
\begin{figure}
	\centering
	\subfigure[Left ventricular bundle brunch block.]{\label{fig:VBB-left}
	\includegraphics[trim = 15mm 0mm 5mm 0mm, clip, width=\textwidth]{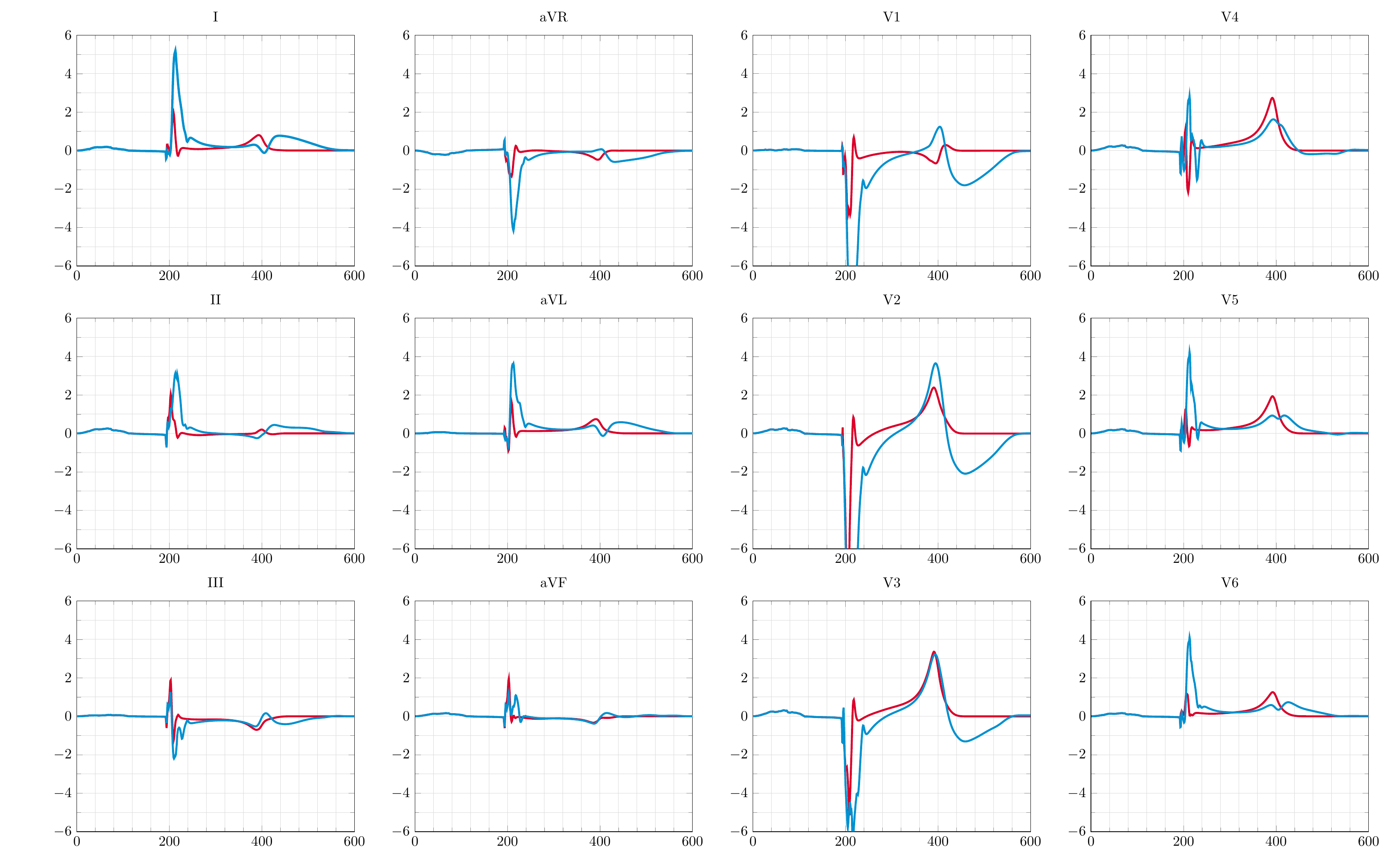}
	}\\
	\subfigure[Right ventricular bundle brunch block.]{\label{fig:VBB-right}
	\includegraphics[trim = 15mm 0mm 5mm 0mm, clip, width=\textwidth]{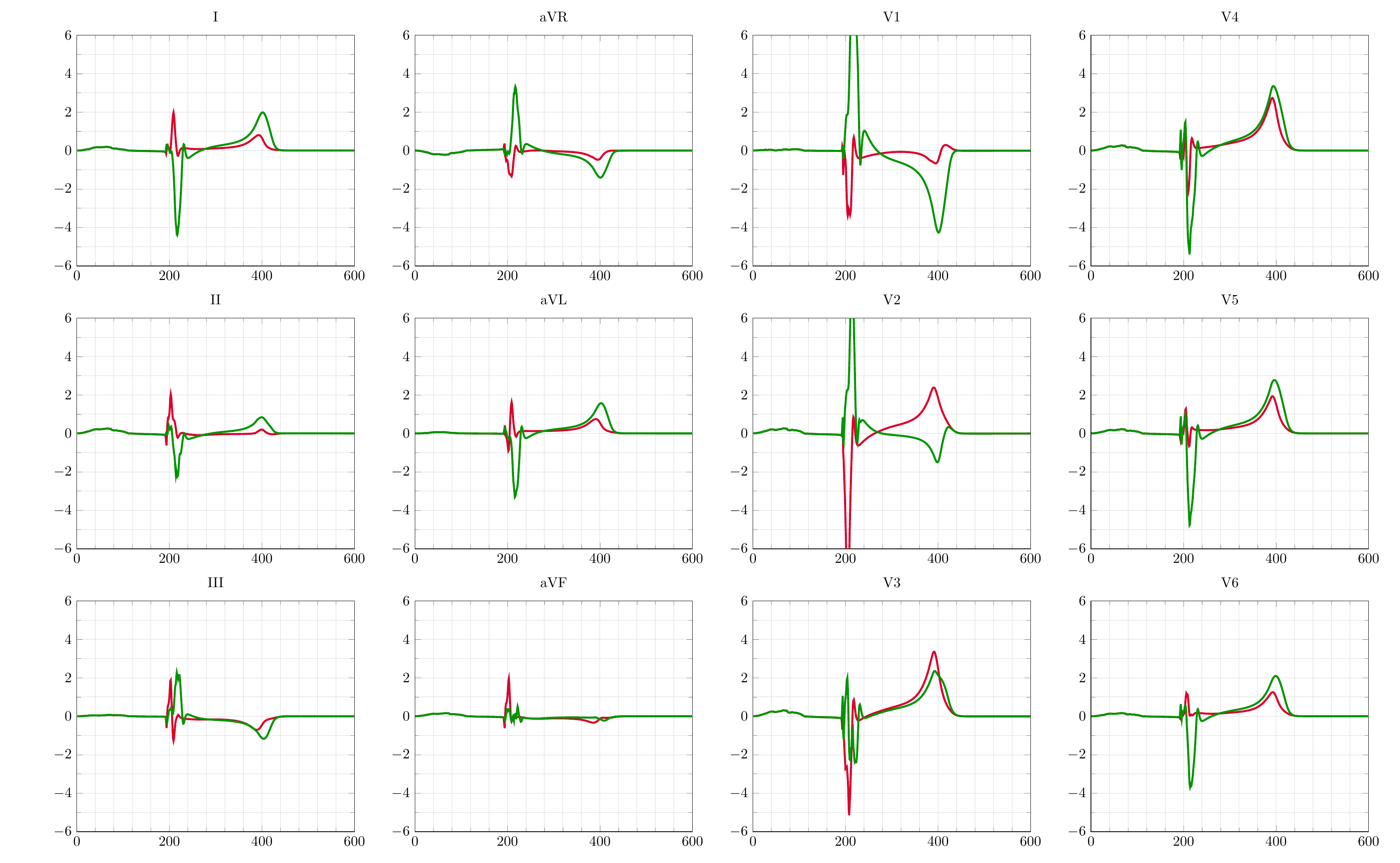}
	}
	\caption{Left and Right Bundle Brunch Block --~Healthy case in \textcolor{myred}{red}, LBBB in \textcolor{myblue}{blue} and RBBB in \textcolor{mygreen}{green} (voltages (mV) versus time (ms))}
	\label{fig:VBB}
\end{figure}

We start with a left or a right bundle branch block. In a healthy case, the right and the left ventricles are activated simultaneously. Now, in order to simulate a left (or a right) bundle branch block, the initial activation is blocked in the left (resp. right) ventricle. On the left of Figure~\ref{fig:pathologies}, we can see a left ventricular block. In order to obtain a left (resp. right) bundle branch block, the depolarization of the left (resp. right) Purkinje fibers is delayed~\cite{boulakia-cazeau-fernandez-10}. Results are reported in Figure~\ref{fig:VBB} for the left and right bundle branch blocks. We recognize the main characteristics reported in the medical literature: larger QRS, lead V1 without Q-wave \cite{malmivuo-plonsey-95}, leads V1 and V6 similar to those presented in Figure~\ref{fig:pathologies} (left). The QRS-complex exceeds 0.1 seconds in both cases. Furthermore, it can be seen in Figure~\ref{fig:VBB} that the duration between the beginning of the QRS complex and its last positive wave in V1 (resp. V6) exceeds 0.04 seconds which is a known sign of right (resp. left) bundle branch block~\cite{malmivuo-plonsey-95}. 

\subsubsection{Bachmann's bundle block}
\begin{figure}
	\centering
	\includegraphics[trim = 15mm 0mm 5mm 0mm, clip, width=\textwidth]{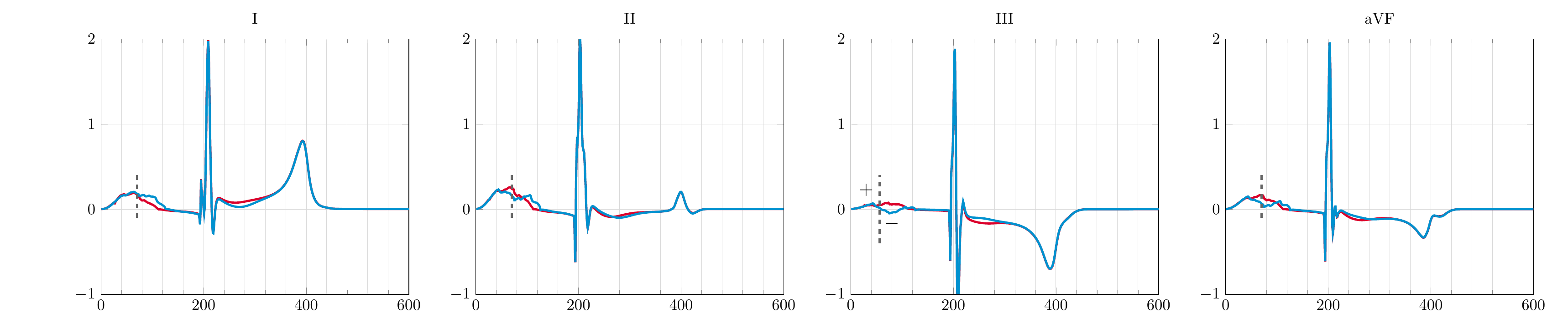}
	\caption{Bachmann's Bundle Block --~Healthy case in \textcolor{myred}{red}, BBB in \textcolor{myblue}{blue} (voltages (mV) versus time (ms))}
	\label{fig:PartialBBB}
\end{figure}

In the heart conduction system, the Bachmann bundle connects the left atrium with the right atrium and is the preferential path for the electrical activation of the left atrium. A block of the Bachmann bundle is represented in the middle of Figure~\ref{fig:pathologies}. It is characterized by the presence of P-wave duration that equals or exceeds 0.12 seconds and presents usually a bimodal morphology, especially in leads I, II, aVF and the lead III becomes biphasic, as we can see in Figure~\ref{fig:pathologies}. This is a very specific sign of left atrial enlargement \cite{malmivuo-plonsey-95,netter-69}. We simulate it by decreasing the maximal conductance $g_{Na} = 7.8$ in the Bachmann bundle. The results are given in Figure~\ref{fig:PartialBBB}. The more important the block, the more negative the P wave on lead III. A negative P wave  in the third lead corresponds to the retrograde depolarization of the left atrium. The morphology of the simulated P wave is in very good agreement with the criteria given in the literature for various degrees of Bachmann's bundle blocks~\cite{bayesdeluna-guindo-vinolas-99,bayesdeluna-platonov-cosio-12}.

%\begin{figure}
%	\centering
%	\includegraphics[trim = 15mm 0mm 2mm 0mm, clip, width=\textwidth]{ecgWenckebach.pdf}
%	\caption{Wenckebach atrioventricular block (voltages (mV) versus time (ms))}
%	\label{fig:SecondAVB}
%\end{figure}

\subsubsection{Wolff-Parkinson-White syndrome}
\begin{figure}
	\centering
	\includegraphics[trim = 15mm 0mm 5mm 0mm, clip, width=\textwidth]{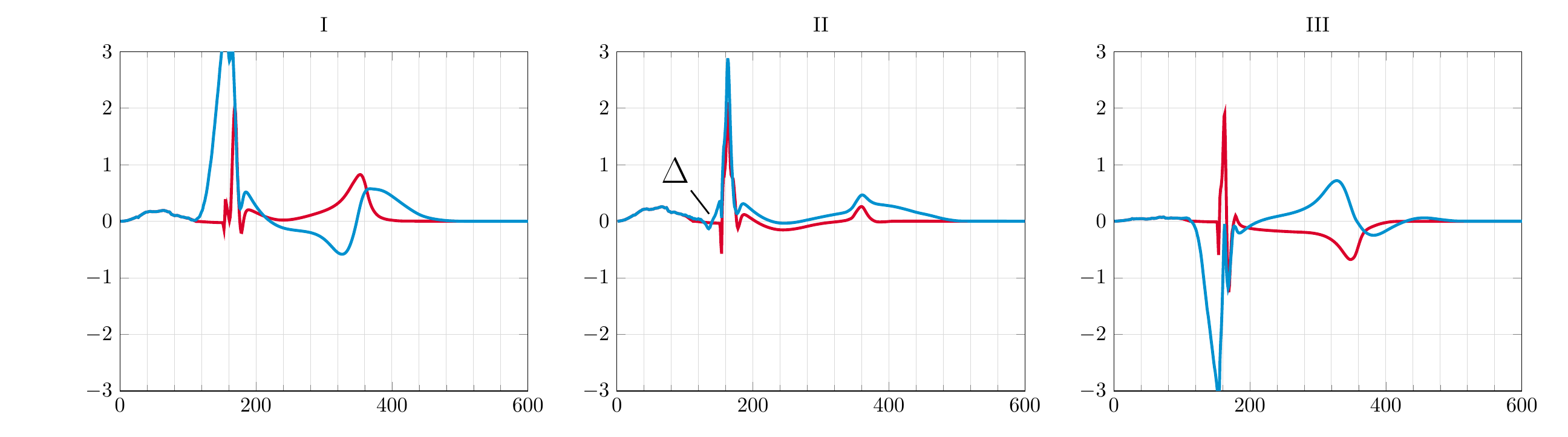}
	\caption{Wolff-Parkinson-White syndrome --~Healthy case in \textcolor{myred}{red} and WPW in \textcolor{myblue}{blue} (voltages (mV) versus time (ms))}
	\label{fig:WPW}
\end{figure}

The Wolff-Parkinson-White syndrome is one of the numerous pathologies of the conduction system of the heart. It corresponds to a pre-excitation syndrome and is caused by the presence of one or more abnormal electrical conduction pathways between the atria and the ventricles, called bundles of Kent. Electrical signals travel down these abnormal pathways and may stimulate the ventricles prematurely. In the right of Figure~\ref{fig:pathologies}, we can see a schematic of the Wolff-Parkinson-White syndrome. Here we consider one Kent bundle only and we model this abnormal pathway by stimulating a ventricle area near of the atria represented at the right of Figure~\ref{fig:connectionSurfaceKB}. The Wolff-Parkinson-White syndrome is commonly diagnosed with the electrocardiogram \cite{rosner-brady-kefer-99}. It is characterized by a delta wave, a slurring of the initial segment of the QRS complex, due to the arrival of the impulses in the ventricles via the abnormal route, which is associated with a short PR interval. Another feature is a QRS complex widening with a total duration greater than 0.12 seconds. We can indeed observe these characteristics, in particular the delta wave in Figure~\ref{fig:WPW}. 

\subsection{Comparison with the Mitchell-Schaeffer model}
In this section we are interested in the impact of the ionic model on the ECG simulation. The membrane current is now described with the Mitchell-Schaeffer model~\cite{mitchell-schaeffer-03} which is a one-current phenomenological ionic model, offering interesting properties with a very limited number of parameters. 

The Mitchell-Schaeffer model is applied with the same conductivity parameters (except for some atrial areas, see below) and the same initial stimulus as described above. Table~\ref{tab:parametersMSModel} gives the value of the Mitchell-Schaeffer parameters, appropriately rescaled~\cite{boulakia-cazeau-fernandez-10}. In order to correctly reproduce the T wave, we take into account three layers of cells in the left ventricles and an homogeneous tissue in the right ventricle, as described in Section~\ref{sec:wholeHeart}. The $\tau_{close}$ parameter varies according to the type of cell (Table~\ref{tab:parametersMSModel}). In the atria, the repolarization ``propagates'' in the same direction as the depolarization. We therefore take a constant value for $\tau_{close}$, equal to $100$ms. As previously explained, we changed the values of the maximal conductance $g_{Na}$ in the different atrial areas in the Courtemanche-Ramirez-Nattel model in order to take into account these bundles. The Mitchell-Schaeffer model does not allow the same flexibility, so we decide to directly modify the value of conductivity parameters. The atrial conductivities are modified as reported in Table~\ref{tab:atriaCondParam} in order to represent the different slow and fast bundles. 

\begin{table}[ht]
\centering
\begin{tabular}{cccccccccc}
\hline
\hline
$\tau_{in}$ & $\tau_{out}$ & $\tau_{open}$ & $\tau_{close}^\text{endo}$ & $\tau_{close}^\text{Mcell}$ & $\tau_{close}^\text{epi}$ & $\tau_{close}^\text{RV}$ & $V_{min}$ & $V_{max}$ &  $V_{gate}$ \\ 
\multicolumn{2}{c}{$(\textrm{cm}^{2}\hspace{-0.03cm}.\textrm{mA}^{\hspace{-0.07cm}-\hspace{-0.03cm}1})$} & \multicolumn{5}{c}{$(\textrm{ms})$} & \multicolumn{3}{c}{$(\textrm{mV})$} \\
\hline
\hline
$4.0$ & $90.0$ & $300.0$ & $120.0$ & $100.0$ & $80.0$ & $90.0$ & $-80.0$ &  $20.$ & $-67.0$ \\
\hline
\end{tabular}
\caption{Mitchell and Schaeffer parameters and constants (different values of $\tau_{close}$ are given because of an heterogeneous tissue is considered~\cite{boulakia-cazeau-fernandez-10})}
\label{tab:parametersMSModel}
\end{table}

\begin{table}[ht]
\centering
\begin{tabular}{cccccccc}
\hline
\hline
& regular tissue & PM & CT & BB & FO\\
\hline
\hline
$\sigma_i^{a,t}$ &  $2.5\, 10^{-4}$ &  $4.5\, 10^{-4}$  &  $7.5\, 10^{-4}$  &  $1.19\, 10^{-3}$ &  $2.5\, 10^{-4}$ \\
$\sigma_e^{a,t}$ &  $9.0\, 10^{-4}$ &  $1.35\, 10^{-3}$ &  $2.7\, 10^{-3}$  &  $4.3\, 10^{-3}$  &  $9.0\, 10^{-4}$ \\
$\sigma_i^{a,l}$ &  $2.5\, 10^{-3}$ &  $4.5\, 10^{-3}$  &  $1.09\, 10^{-2}$ &  $1.86\, 10^{-2}$ &  $2.27\, 10^{-3}$\\
$\sigma_e^{a,l}$ &  $2.5\, 10^{-3}$ &  $4.5\, 10^{-3}$  &  $1.09\, 10^{-2}$ &  $1.86\, 10^{-2}$ &  $2.27\, 10^{-3}$\\
\hline
\end{tabular}
\caption{Atrial conductivity parameters (all in $\textrm{S.cm}^{-1}$) for the Mitchell-Schaeffer model}
\label{tab:atriaCondParam}
\end{table}

\begin{figure}[ht!]
	\centering
	\includegraphics[trim = 15mm 0mm 5mm 0mm, clip, width=\textwidth]{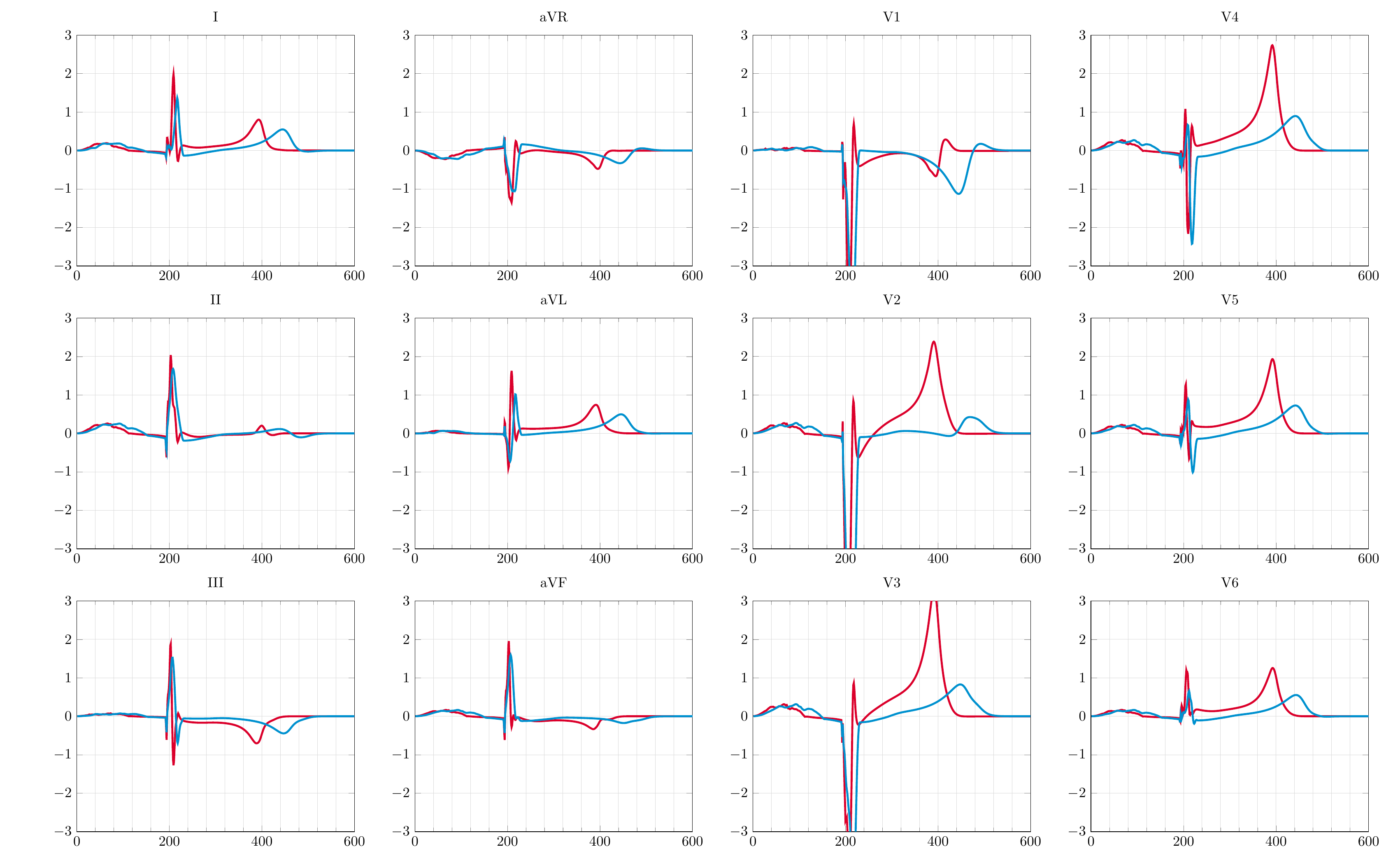}
	\caption{ECGs obtained with different ionic models --~Courtemanche/MV ionic model in \textcolor{myred}{red} and Mitchell and Schaeffer model in \textcolor{myblue}{blue} (voltages (mV) versus time (ms))}
	\label{fig:MS}
\end{figure}

Figure~\ref{fig:MS} shows the ECGs obtained with the Mitchell-Schaeffer (MS) model and the combined Courtemanche/MV model. We can see that the results are reasonably close. With the MS model some oscillations in the P wave and the QRS complex are fixed, but the R wave progression in precordial leads is less satisfactory and the T wave in V2 and V3 are not correct. In spite of these flaws, those results are reasonable, and could probably be improved with a finer choice of the parameters. It is interesting to note that the results of the simulations are robust with respect to the choice of the ionic model: the Courtemanche/MV model in general gives better results, but it can be replaced by the MS model in order to reduce the computational costs and the model complexity without affecting too much the ECG.  This remark is especially important if the ECG simulator has to be used for inverse problems: in that case, a model with a reduced number of parameters may be more attractive. 

%=========================================================================

\section{Virtual ``electrode vest''} \label{sec:grid}

\begin{figure}[ht]
	\centering
	\includegraphics[width=\textwidth]{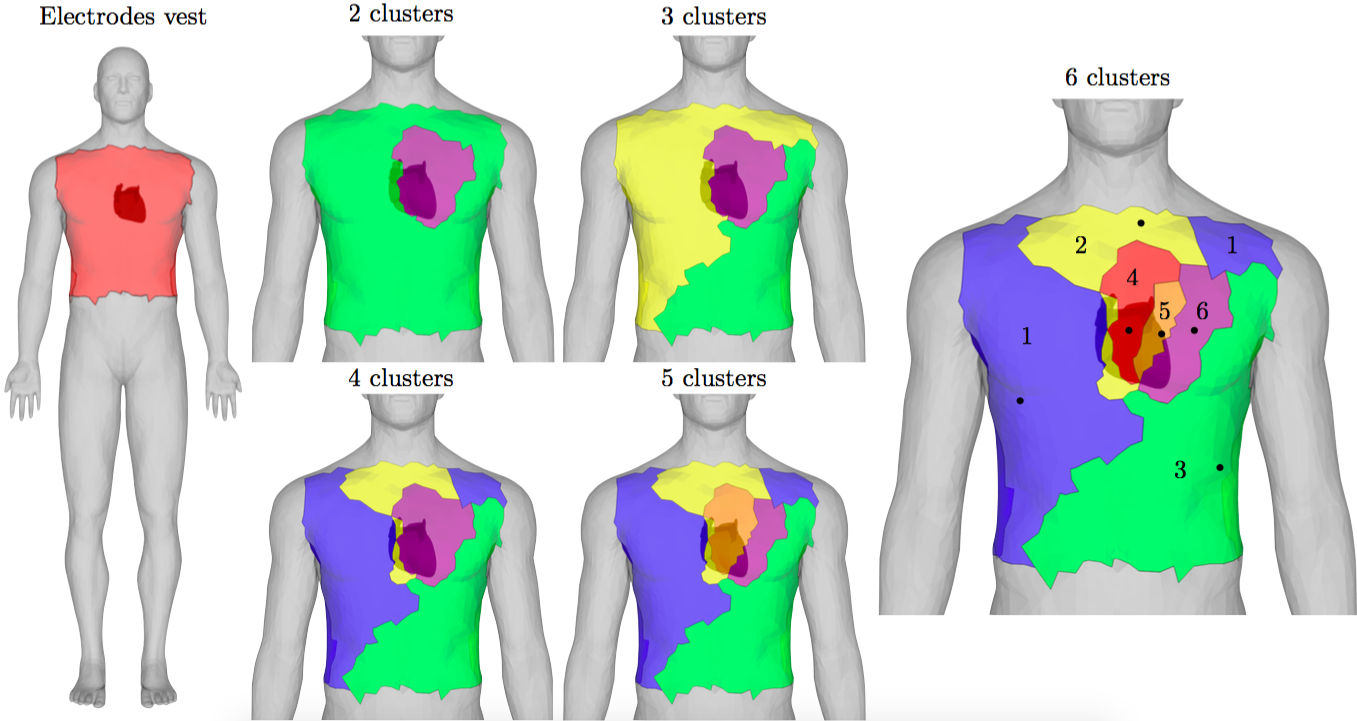}
	\caption{Human body mesh, ``electrode vest''(left) and clustering results (right). From left-top to right-bottom clustering agglomeration, from 2 to 6 clusters.}
	\label{fig:body-mesh-cluster}
\end{figure}

In the previous sections, we focused on the 12-lead ECG because it is widely used in practice, and it can be easily assessed with the medical literature. But our simulator can of course provide more sophisticated measurements, like those obtained with electrode vests. 

\begin{figure}[ht]
	\centering
	\includegraphics[width=0.7\textwidth]{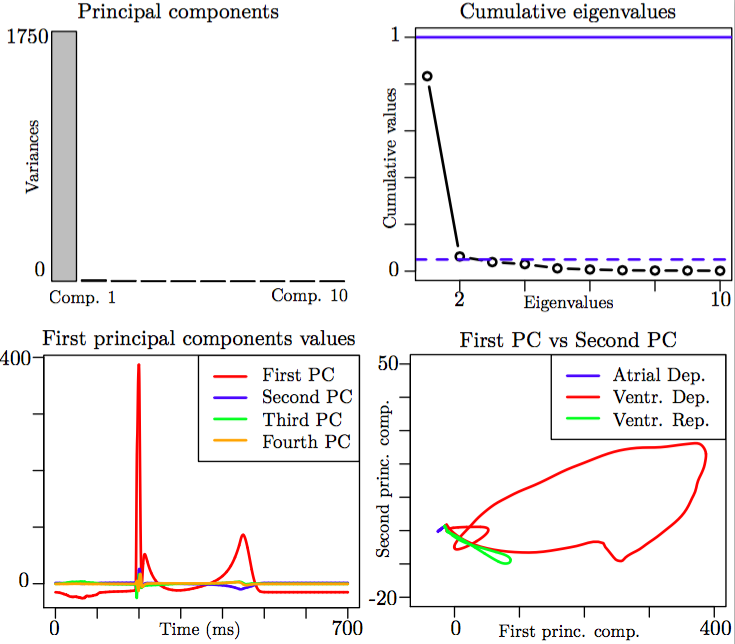}
	\caption{Principal components analysis on electrode vest signals}
	\label{fig:ecg-pc}
\end{figure}

Many studies have been carried out on this topic: on the forward problem and the analysis of the number of  electrodes~\cite{huiskamp-greensite-97,hoekema-uijen-vanOosterom-99}, or on the reconstruction of the potential on the heart surface~\cite{pullan-buist-cheng-05,sundnes-06}. Our objective here is less ambitious: we just show an example of a statistical analysis that can be done with the measures provided by our ECG simulator. 

To do so, we simulate a virtual ``electrode vest'' which contains $N_{ECG} = 1,216$ electrodes. Figure~\ref{fig:body-mesh-cluster}~(left) shows the ``electrode'' locations, which correspond to all the nodes of the mesh in the red region of the torso. The heart geometry used in this study contains $N_{\partial \Omega_H} = 28,510$ boundary vertices. We compute the body surface potential as described in~Section~\ref{sec:modeling_assump}.

\begin{figure}[ht!]
	\centering
	\includegraphics[width=\textwidth]{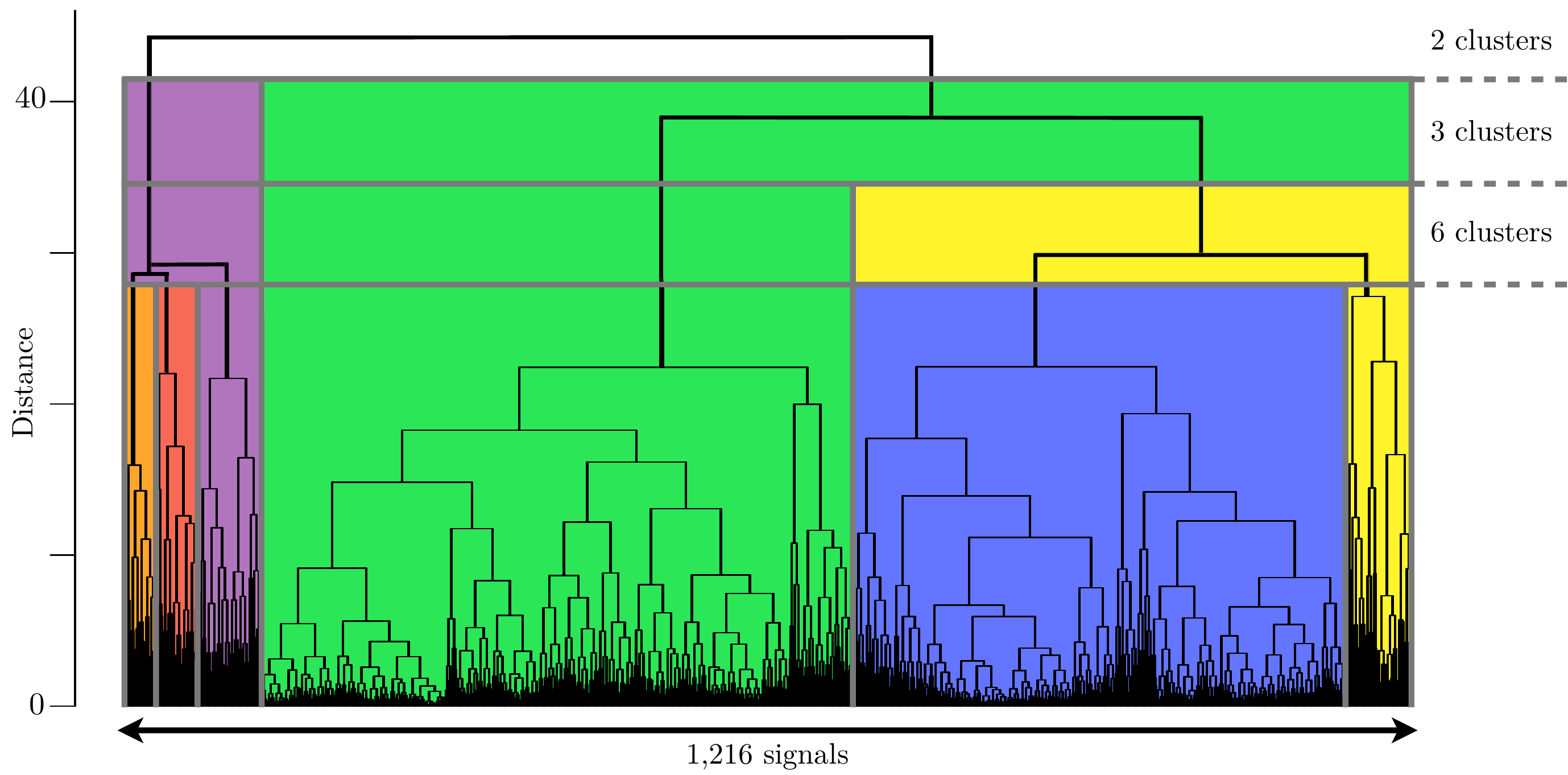}
	\caption{Clustering agglomeration of electrodes measures potential using euclidean distance and complete linkage. Colors represent the clusters of points as shown in Figure~{fig:body-mesh-cluster}.}
	\label{fig:ecg-clust-box}
\end{figure}

We are interested in analyzing the electrode signals with respect to their positions on the body. In order to divide the signals into different groups, an agglomerative hierarchical clustering analysis is applied to the $1,216$ measures. Clustering is a statistical technique used to classify data based on \emph{similarities}, or \emph{distances}, between them. In particular, in agglomerative hierarchical clustering method, at the beginning each individual (data) represents a group itself. Then, these groups are merged together according to their decreasing distance. The procedure is schematized in Figure~\ref{fig:ecg-clust-box}: the bottom of the clustering tree represents the $1,216$ signals, the top represents a unique group. In order to measure the similarity between data, an Euclidean distance is used
\begin{equation}
\text{dist}(V_i,V_j) = \sqrt{\sum_n \big(V_i(t^n)-V_j(t^n) \big)^2}
\end{equation}
where $V_i$, $i=1,\ldots,1216$, represents an electrode signal. The criterion used to merge groups (\emph{linkage}) is based on the distance between two groups, defined as the maximal distance between the data of the first group and the data of the second one
\begin{equation}
\text{dist}(C_i,C_j) = \max_{k,h} \big(\text{dist}(\{V_k \in C_i \}, \{ V_h \in C_j \}) \big)
\end{equation}
where $C_i$ represents the $i-$th cluster~\cite{johnson-wichern-07}.

\begin{figure}[ht!]
	\centering
	\includegraphics[width=0.8\textwidth]{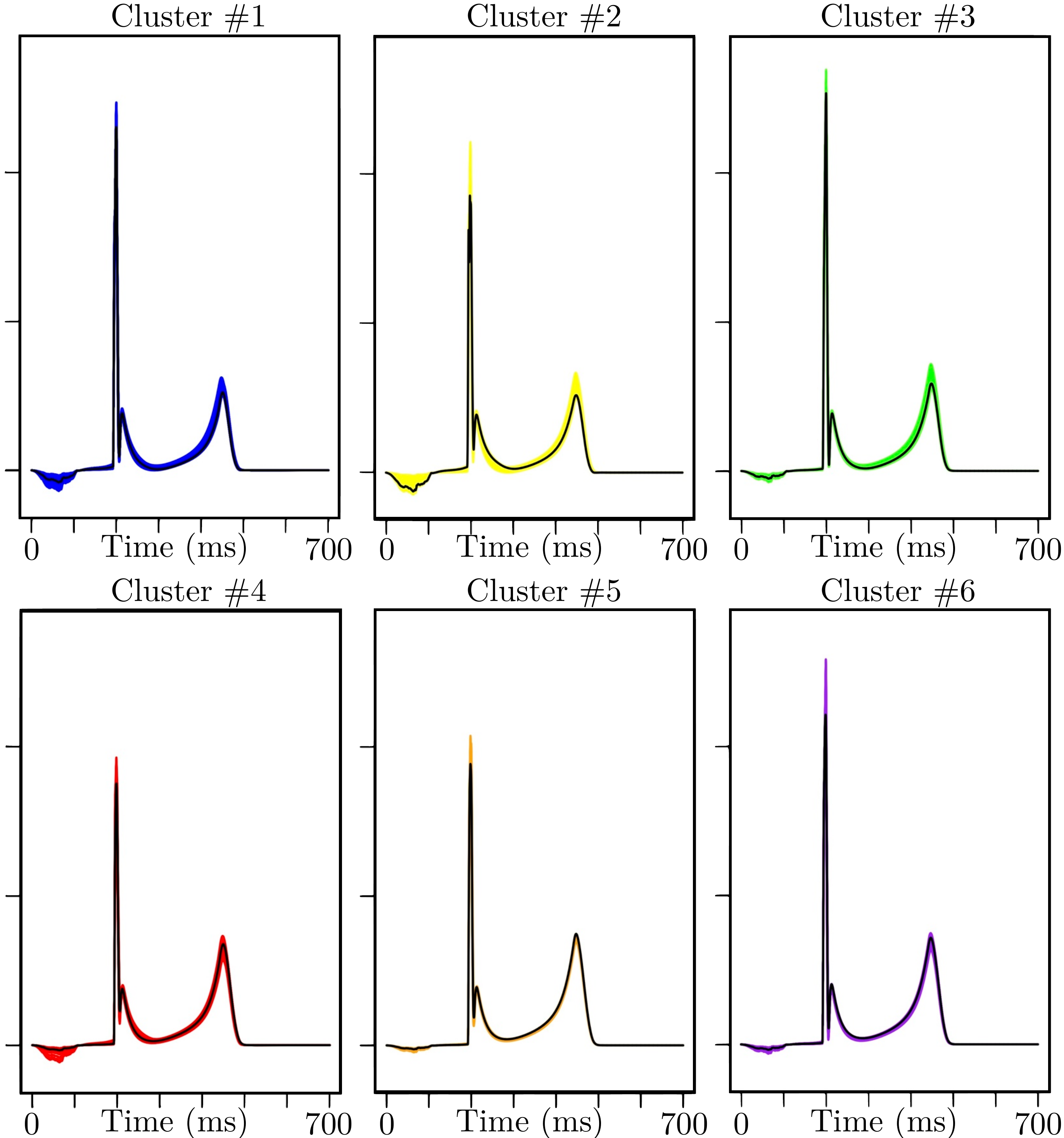}
	\caption{Signals in different clusters as shown in Figure~\ref{fig:body-mesh-cluster}(right). Black lines indicate the ``mean'' point (points of Figure~\ref{fig:body-mesh-cluster}(right))}
	\label{fig:ecg-clust-mes}
\end{figure}

In Figure~\ref{fig:body-mesh-cluster} we show the agglomeration procedure from 2 to 6 clusters. First, the division into two clusters indicates a separation between the heart region (purple area of Figure~\ref{fig:body-mesh-cluster}) and the rest of the vest (green region). Second, a division into 2 parts underlies the atria-ventricular axes (separation of green and yellow zones). Then, we can see from the clustering agglomeration plot of Figure~\ref{fig:ecg-clust-box}, that the separations into 4, 5 and 6 groups (blue, orange and red areas) are made at very closed distances. The last subdivision we consider is made of 6 groups. In Figure~\ref{fig:ecg-clust-mes} we plot the signals belonging to the 6 clusters and their center values. The center values of each cluster are computed as the point minimizing the distance between itself and the other points of the cluster
\begin{equation}
m_i = \arg \min_j \big( \sum_k \text{dist}(V_j,V_k)  \big), \ \ \forall i.
\end{equation}

\begin{figure}[ht!]
	\centering
	\includegraphics[width=0.7\textwidth]{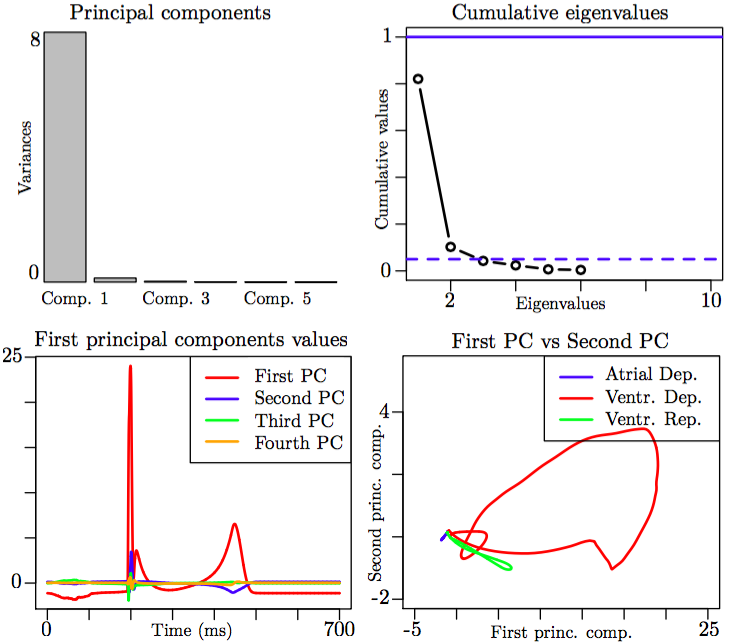}
	\caption{Principal components analysis on 6 electrodes signals.}
	\label{fig:ecg-cluster}
\end{figure}

Then, we re-apply the principal components decomposition on the 6 centers of the cluster, the points indicated in Figure~\ref{fig:body-mesh-cluster} (right). Comparing results of Figure~\ref{fig:ecg-cluster} with Figure~\ref{fig:ecg-pc} we observe that the first principal component is again much larger than the other ones, and the same curves represent the first, second and third principal components.

The two approaches -- the principal components decomposition and the hierarchical clustering analysis -- suggest that it is not necessary to have a very high number of skin electrodes to describe the body surface potential. A limited number of correctly positioned electrodes should be enough to represent most of the features of the signal. A deeper analysis of the number and the locations of the electrode goes beyond the scope of this study. The purpose here was just to illustrate that our simulator could be used to generate more general signals than the standard 12-lead ECGs. 
 
\section{Limitations}
\label{sec:limitations}
This work can be considered as a step forward towards the simulation of ECGs with a biophysical model. However, it requires further improvement and several limitations should be mentioned. 

A strong coupling with the torso should be done to assess the impact of the isolated heart assumption \eqref{eq:isolated}. It was indeed shown in \cite{boulakia-cazeau-fernandez-10} that this assumption modified the amplitude of the QRS with a factor two without affecting the shape significantly. But it is possible that the isolated heart assumption has a stronger effect with other activation sequences. 

The role of parameter $R_p$ in~\eqref{eq:resistcapaccouplWeak} is not completely understood. To get correct relative amplitudes of the P and R waves, we took {\em ad hoc} values, higher in the atria than in the ventricle. We postulated that this might be due to a difference of conductivity at the interfaces ventricle-torso and atria-torso but we did not find any evidence in the literature to support this assumption. This choice should therefore be seen as a research hypothesis which has to be further investigated.

Another limitation that should be mentioned is the rough modeling of the atrio-ventricular node and the Purkinje fibers. 

At last, by adjusting some parameters of the models, it is likely that the realism of the simulated ECGs could still be improved. This task, which was mainly done manually in this work, should be addressed with an inverse problem strategy \cite{boulakia-schenone-gerbeau-12,corrado:hal-01091751}.

\section{Conclusions}\label{sec:conclusions}
\label{sec:conclusion}

We have presented a comprehensive model for the simulation of full cycle ECGs. The main ingredients are: volume bidomain equations for the ventricular part coupled with the MV ionic model; surface bidomain equations for the atrial part coupled with the Courtemanche ionic model; one-way coupling between the heart and the torso through a resistor transmission condition.

This model has provided a healthy ECG whose qualities and flaws have been assessed with several qualitative and quantitative criteria extracted from the medical literature. Four pathological cases have been investigated: left and right bundle branch blocks, Bachmann's bundle block,  and Wolff-Parkinson-White syndrome. In the healthy case, we have shown that similar results can be obtained with the one-current phenomenological Mitchell-Schaeffer model. The simulation of a virtual vest of electrodes has illustrated a potential application of this simulator.  

As a natural continuation of this study, the proposed electrical model of the heart could also be used in electromechanical simulations, as was done in \cite{chapelle-fernandez-zemzemi-09,corrado:hal-01091751}, including a mechanical model of the atria.

\newpage
\newpage

\bibliographystyle{plain}
\bibliography{biblio} 

\end{document}